\documentclass{aa}
\usepackage{graphicx}
\usepackage{txfonts}

\usepackage{natbib}
\bibpunct{(}{)}{;}{a}{}{,} 
\usepackage{color}
\usepackage[colorlinks=true,linkcolor=black,citecolor=blue,urlcolor=blue]{hyperref}

\makeatletter
\renewcommand*\aa@pageof{, page \thepage{} of \pageref*{LastPage}}
\makeatother

\newcommand{\prodimo}{P{\tiny RO}D{\tiny I}M{\tiny O}\;}
\newcommand{\prodimonospace}{P{\tiny RO}D{\tiny I}M{\tiny O}}
\newcommand{\oiline}{$\mathrm{[OI]}\,0.63\,\mathrm{\mu m}\,$}
\newcommand{\htline}{$\mathrm{o\!-\!H_2}\,2.12\,\mathrm{\mu m}\,$}

\begin{document} 

  \title{Interpreting molecular hydrogen and atomic oxygen line emission of \mbox{T Tauri} disks with photoevaporative disk-wind models.}
  \author{Ch. Rab\inst{1,2} 
  \and M. Weber\inst{1,3}
  \and T. Grassi\inst{2,1,3}
  \and B. Ercolano\inst{1,3}
  \and G. Picogna\inst{1,3}
  \and P. Caselli\inst{2}
  \and W.-F. Thi\inst{2}
  \and I. Kamp\inst{4}
  \and P. Woitke\inst{5}
  }
  
  \institute{University Observatory, Faculty of Physics, Ludwig-Maximilians-Universität München, Scheinerstr. 1, 81679 Munich, Germany \email{rab@mpe.mpg.de} 
  \and Max-Planck-Institut f\"ur extraterrestrische Physik, Giessenbachstrasse 1, 85748 Garching, Germany 
  \and Excellence Cluster Origin and Structure of the Universe, Boltzmannstr.2, D-85748 Garching bei M{\"u}nchen, Germany
  \and Kapteyn Astronomical Institute, University of Groningen, P.O. Box 800, 9700 AV Groningen, The Netherlands 
  \and Space Research Institute, Austrian Academy of Sciences, Schmiedlstr. 6, A-8042, Graz, Austria
  } 
   \date{Received June 23 2022 / Accepted Oct 25 2022}
\abstract
{Winds in protoplanetary disks play an important role in their evolution and dispersal. However, the physical process that is actually driving the winds is still unclear (i.e. magnetically versus thermally driven), and can only be understood by directly confronting theoretical models with observational data. 
}
{We aim to interpret observational data for molecular hydrogen and atomic oxygen lines that show kinematic disk-wind signatures in order to investigate whether or not purely thermally driven winds are consistent with the data.}
{We use hydrodynamic photoevaporative disk-wind models and post-process them with a thermochemical model to produce synthetic observables for the spectral lines $\mathrm{o\!-\!H_2}\,1\!-\!0\,\mathrm{S}(1)$ at $2.12\;\mathrm{\mu m}$ and $\mathrm{[OI]}\,  ^1\mathrm{D}_2\!-\!{^3}\mathrm{P}_2$ at $0.63\,\mathrm{\mu m}$  and directly compare the results to a sample of observations.}
{We find that our photoevaporative disk-wind model is consistent with the observed signatures of the blueshifted narrow low-velocity component (NLVC) ---which is usually associated with slow disk winds--- for both tracers. Only for one out of seven targets that show blueshifted NLVCs does the photoevaporative model fail to explain the observed line kinematics. Our results also indicate that interpreting spectral line profiles using simple methods, such as the thin-disk approximation, to determine the line emitting region is not appropriate for the majority of cases and can yield misleading conclusions. This is due to the complexity of the line excitation, wind dynamics, and the impact of the actual physical location of the line-emitting regions on the line profiles.} 
{The photoevaporative disk-wind models are largely consistent with the studied observational data set, but it is not possible to clearly discriminate between different wind-driving mechanisms. Further improvements to the models  are necessary, such as consistent modelling of the dynamics and chemistry, and detailed modelling of individual targets (i.e. disk structure) would be beneficial. Furthermore, a direct comparison of magnetically driven disk-wind models to the observational data set is necessary in order to determine whether or not spatially unresolved observations of multiple wind tracers are sufficient to discriminate between theoretical models.}
\titlerunning{Modelling molecular and atomic disk wind tracers.}
\maketitle
\section{Introduction}
Understanding the dispersal and evolution of protoplanetary disks, the birthplaces of planets, is crucial to understanding the planet-formation process. Disk winds, either thermally or magnetically driven, likely play a crucial role in the disk mass loss and ---in the case of magnetically driven winds--- angular momentum extraction. Although several theoretical models exist to understand the physics of disk winds and their impact on disk evolution \citep[see reviews of][]{Gorti2016,Ercolano2017,Pascucci2022,Lesur2022}, questions remain as to: whether magnetohydrodynamic (MHD) or thermally driven photoevaporative winds dominate in disks; the nature of their interplay; and whether or not the outflows evolve from being MHD-dominated in young objects to thermally dominated at later ages \citep[e.g.][]{Ercolano2017}. Observationally constrained theoretical models are necessary to tackle these questions. Specifically, wind dynamics models are necessary as an input to astrochemical models (especially for molecular tracers) in order to obtain synthetic observables.

The most commonly observed tracers of outflows and winds in the protoplanetary disk stage are forbidden lines of atoms and low-ionisation species such as [OI], [SII], [NII], [FeII], and [NeII], with the most prominent being the \oiline \citep[e.g.][]{BaldovinSaavedra2012,Rigliaco2013,Simon2016,Fang2018,Banzatti2019}. High-spectral-resolution observations ($\Delta \varv\!\approx4\!-\!7\,\mathrm{km s^{-1}}$) of the \oiline commonly show complex line profiles that can be decomposed into several Gaussian components. The two main components are a high-velocity component (HVC), showing a shift of the peak flux location of $\varv_\mathrm{p}\!<\!-30\,\mathrm{km s^{-1}}$ (relative to the systemic velocity) and a low-velocity component (LVC) showing blueshifted emission with $\varv_\mathrm{p}\!>\!-30\,\mathrm{km s^{-1}}$. The HVC is often attributed to jets and strong outflows, whereas the LVC is supposed to trace slower disk winds. The LVC is also often further decomposed into two Gaussian components, a broad LVC (BLVC; i.e. full width at half maximum \mbox{$>30\,\mathrm{kms^{-1}}$}) that dominates the line wings and a narrow LVC (NLVC) with a blueshifted peak of a few $\mathrm{km s^{-1}}$, but single Gaussian fits to LVCs seem to be as common \citep[see e.g.][]{Banzatti2019}.

Alternatives to emission lines are absorption lines, such as $\mathrm{CII}\,\lambda 1335$\AA, which was~used by \citet{Xu2021} to study winds and outflows in T Tauri stars. These authors find that their findings can be best explained by a combination of magnetically and thermally driven flows. Molecular tracers of slow disk winds observed at high spectral resolution are less common and are mostly restricted to CO \citep[e.g.][]{Pontoppidan2011,Klaassen2013,Banzatti2022} and $\mathrm{H_2}$ \citep[e.g.][]{Beck2008,Beck2019}. However, molecular tracers are of particular interest as they can trace different physical regions of the wind than the atomic tracers and provide further constraints for measuring mass-outflow rates.

A consistent physical interpretation of all the available observational data is challenging and often the different tracers are interpreted individually, in particular as simultaneous observations of multiple tracers come with significant difficulty. However, \citet{Gangi2020} presented new data for the \oiline and the \mbox{ortho-H$_2$} (hereafter o-H$_2$) $2.12\,\mathrm{\mu m}$ spectral lines for a sample of $36$ young disk-bearing stars. The data set of these latter authors is unique as both lines were observed simultaneously, and for seven targets, where both lines are detected, kinematic signatures of winds for both tracers were identified. \citet{Gangi2020} interpreted their data as a potential indication of magnetically driven winds, as their analysis indicates that both spectral lines originate from similar regions of the disk wind. Also, several of the previously mentioned observational studies postulate that MHD-driven winds \citep[e.g.][]{Fang2018,Whelan2021} or a combination of MHD- and thermally driven winds \citep[e.g.][]{Xu2021} are common in T Tauri stars. The wealth of observational data provides strong constraints for theoretical models. However, a direct comparison of the observational data to the model is challenging, as it requires detailed models for the disk dynamics, evolution, thermal structure, radiation environment, and chemistry, as well as proper modelling of synthetic observables.

Several disk-wind modelling approaches exist. \cite{Ercolano2008c} and \citet{Gorti2009} presented static photoevaporative disk-wind models with a focus on radiative transfer and ionisation physics, with their main aim being to interpret atomic wind tracers. The first models that coupled X-ray and extreme ultraviolet (EUV) radiation physics to hydrodynamic models for photoevaporative winds were performed by \citet{Owen2010} and more recently by \citet{Picogna2019,Picogna2021}. By applying a post-processing step, those models can be used to predict atomic line emission \citep[e.g.][]{Ercolano2016} or for direct comparison to observations \citep[e.g.][]{Weber2020}. In addition to the photoevaporative models, \citet{Weber2020} used semi-analytic MHD wind models and suggested that the multi-component [OI] spectral line profiles could be explained by a combination of thermally driven (LVC) and MHD-driven (HVC) wind components. 

More consistent wind models, coupling the dynamics, radiation physics, and thermochemistry for photoevaporative winds, are presented by \citet{Wang2017a} and \citet{Nakatani2018a,Nakatani2018}, who show that molecular hydrogen can survive in several of their models (i.e. depending on the included  heating sources). The first theoretical models for MHD outflows and winds that include chemistry were presented by \citet{Panoglou2012}. These authors follow the chemical evolution along individual wind stream lines, and one of their main findings is that, even in the T Tauri stage, molecular hydrogen can survive in the wind regions. Full MHD disk-wind models including thermochemistry were built by \citet{Wang2019} and \citet{Gressel2020}. These latter models indicate that both magnetic and thermal effects (i.e. radiation) can play an important role in slow disk winds. Such self-consistent disk-wind models are computationally expensive and therefore a direct comparison to observations is difficult; only \citet{Gressel2020} presented synthetic observables for their models, but these authors focused on far-infrared (FIR) and sub-millimetre (submm) line emission. 

In this work, we aim to interpret observations of spectral lines  \mbox{$\mathrm{o\!-\!H_2}\,1\!-\!0\,\mathrm{S(1)}$} at  $2.12\,\mathrm{\mu m}$ and the $\mathrm{[OI]}\,^1\mathrm{D}_2\!-\!{^3}\mathrm{P}_2$ at $0.63\,\mathrm{\mu m}$  in the context of photoevaporative disk-wind models. Following the approach of \citet{Weber2020} for modelling the atomic line tracers, we use existing radiation-hydrodynamic photoevaporative disk-wind models and post-process them with a radiation thermo-chemical disk code that includes molecular chemistry. This allows us to model the molecular and atomic line emission  simultaneously for \oiline and \htline. With this efficient approach, we aim to interpret the observations of \citet{Gangi2020}, focusing on the line kinematics and the physical origin of the line emission.

We first explain our modelling approach in Sect.~\ref{sec:method}. In Sect.~\ref{sec:results}, we present our results and compare the models to the observational data with a focus on line kinematics. In Sect. 4, we discuss our interpretation of the line kinematics in the context of our models, and simpler approaches and possible next steps for theoretical disk-wind modelling. In Sect.~\ref{sec:conclusions}, we summarise our results and main conclusions.
\section{Method}
\label{sec:method}
\subsection{Radiation thermochemical modelling}
For the physical disk--wind density structure and the velocity field we use the 2D EUV/X-ray (XUV) photoevaporative disk-wind models presented in \citet{Weber2020}. For these models, we use the PLUTO \citep{Mignone2007} hydrodynamic code, extended with a parameterised treatment for the gas temperature \citep{Picogna2019}. The parameterised treatment is based on detailed gas photoionisation and radiative-transfer disk models built with \mbox{MOCASSIN} (e.g. \citealt{Ercolano2003a,Ercolano2005,Ercolano2008}). As shown in \citet{Weber2020}, those models are in good agreement with observed kinematic signatures of the \oiline spectral line; however, they do not include the chemical modelling of molecules such as molecular hydrogen. Therefore, in addition to these models, we use the radiation thermochemical disk code \prodimo (PROtoplanetary DIsk MOdel\footnote{\url{https://prodimo.iwf.oeaw.ac.at} \mbox{Version: 1.0 38955520}}, \citealt{Woitke2009a,Kamp2010,Thi2011,Woitke2016}). \prodimo uses wavelength-dependent continuum radiative transfer to calculate the dust temperature and the radiation field in the disk. Furthermore, \prodimo solves consistently for the gas temperature and the chemical abundances and has a module to produce synthetic observables such as spectral line profiles.

\begin{figure}    
    \includegraphics[width=\hsize]{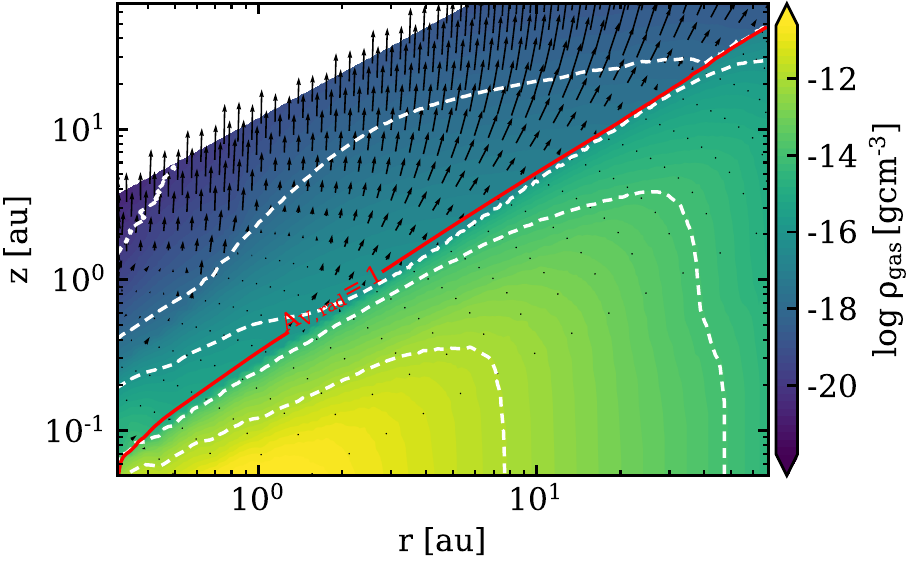}    
    \caption{Two-dimensional disk and wind gas density structure and velocity field (black arrows). This structure is the same in all the presented models. The red solid contour shows where the radial visual extinction $A_\mathrm{V,rad}$ is equal to unity; here, we refer to this layer as the disk surface. The white dashed contours correspond to the values shown in the colour bar.}
    \label{fig:gasstruc}
\end{figure}
For this work, we added an interface to \prodimo that allows us to use the 2D gas density and velocity structure from hydrodynamic models as input.  Figure~\ref{fig:gasstruc} shows the gas density and velocity structure for the disk and wind component as used in \prodimonospace. This structure remains the same for all the models presented in this work (i.e. as the X-ray luminosity is always the same). For the dust structure, we adopt the same dust properties as \citet{Weber2020} and \citet{Ercolano2008c}; an Mathis, Rumpl, Nordsieck (MRN) dust size distribution \citep{Mathis1977} and no settling (see also Sect.~\ref{sec:metmodelseries} and Fig.~\ref{fig:dust}). Besides this interface, it was not necessary to add any other new physical or chemical processes to the \prodimo model. To validate the approach, we compared the results from \prodimo to the results of \citet{Weber2020}, who used  MOCASSIN for post-processing, and find reasonable agreement for both the thermal structure and the synthetic \oiline spectral line predictions. The details of this comparison are described in Appendix\,\ref{sec:moccomp}. We note that we did not attempt to benchmark \prodimo and MOCASSIN, but simply compared the results of both codes for the same physical disk+wind structure. Although there are certain differences in the results, the general agreement is quite remarkable, considering that both codes were developed independently and use different methods for the radiative transfer and thermal-balance modelling for example.

We note that we neglect dynamics when modelling the chemistry, which means that we likely underestimate the amount of molecular hydrogen that might be transported from the disk into the wind region. However, as we model a T Tauri star with a significant far-ultraviolet (FUV) radiation field due to ongoing accretion, it is unlikely that the molecular hydrogen survives for long periods in the disk wind (see also \citealt{Panoglou2012}). Nevertheless, because of this simplifying assumption, the results for the \htline emission likely represent lower limits, as only the molecular hydrogen that is formed and survives in the wind region is modelled (for more details see Sect.~\ref{sec:lineorig}). In future work, we will consider the dynamics and thermochemical processes consistently. However, such models are complex and computational expensive. The approach used in the present work is therefore a first step towards more self-consistent models and allows efficient direct comparison of existing hydrodynamic models to observations (see also Sect.~\ref{sec:disfuture}).
\subsection{Chemical modelling}
For the chemical modelling, we use a chemical network with 100 gas and ice phase species and 1288 chemical reactions, including gas-phase chemistry (i.e. photo-reactions); freeze-out of atoms and molecules; photo-, thermal-, and cosmic-ray desorption of ices; \mbox{X-ray} chemistry \citep[see][]{Aresu2011,Meijerink2012a}; H$_2$ formation on grains; and excited H$_2$ chemistry. This chemical network is described in detail in \citet{Kamp2017}, and was used for example to model multi-wavelength observational data of protoplanetary disks (DIsc ANAlysis project, \mbox{DIANA}\footnote{\url{https://diana.iwf.oeaw.ac.at}}). Also, to the network of \citet{Kamp2017}, we add collisional ionisation of H by electrons using the data from \citet{Janev1993}.

For this work, we focus on molecular hydrogen and therefore summarise the main chemical processes for the $\mathrm{H_2}$ chemistry, in particular the formation and destruction pathways used in our model. We note that \prodimo is a flexible code and other options and chemical rates can be used. Here, we only describe the options adopted in this work. 

For $\mathrm{H_2}$ formation on dust grains, we follow the analytical model of \citet{Cazaux2002,Cazaux2004,Cazaux2010}; furthermore, we include gas-phase formation of $\mathrm{H_2}$ via $\mathrm{H^+}$ and $\mathrm{H{^-}}$ \citep[for details see][]{Woitke2009a,Thi2020a}.
The main $\mathrm{H_2}$ destruction reaction relevant in the wind region is photo-dissociation of $\mathrm{H_2}$. The rate is calculated using the frequency-dependent FUV radiation field from the radiative transfer modelling and detailed photo-dissociation cross-sections \citep{Allison1969, vanDishoeck1988} and by correcting for self-shielding using the approximation of \citet{Draine1996b} (for details see \citealt{Woitke2009a}). We also tested the $\mathrm{H_2}$ self-shielding function of \citet{WolcottGreen2011}, but having found no significant impact on our results (i.e. line profiles), we decided to remain with the prescription of \citet{Draine1996b}. Furthermore, we include excited molecular hydrogen chemistry as described in \citet{Kamp2017}. All other gas-phase chemistry reactions involving $\mathrm{H_2}$ are taken from the UMIST 2012 Database \citep{McElroy2013b}.
\subsection{Line excitation and synthetic observables}
\label{sec:metlines}
For the calculation of the non-local thermodynamic equilibrium (NLTE) level populations and the line cooling for $\mathrm{O}$ and $\mathrm{H_2}$, the escape probability method, including UV pumping, is used in \prodimo \citep{Woitke2009a}. Also included are  fluorescent UV pumping and chemical pumping by OH photo-dissociation for $\mathrm{O}$ \citep{Acke2005}, and pumping by its formation on dust grains for $\mathrm{H_2}$. The details of those pumping mechanisms are described in detail in Appendix A of \citet[][]{Woitke2011}.
  
The collisional and radiative data for $\mathrm{O}$ are from the National Institute of Standards and Technology (NIST) atomic spectroscopic database \citep{Ralchenko2009}, the Leiden Atomic and Molecular Database \citep[LAMDA][]{Schoeier2005c}, \citet{Krems2006}, and \citet{Stoerzer2000}; including collisions with $\mathrm{o\!-\!H_2,p\!-\!H_2,H,H^+}$, and $\mathrm{e^{-}}$. For $\mathrm{H_2}$, the data are from \citet{Wolniewicz1998}, \citet{Wrathmall2007}, and \citet{LeBourlot1999}, including collisions with $\mathrm{o\!-\!H_2,p\!-\!H_2,H,}$ and $\mathrm{He}$. The $\mathrm{H_2}$ ortho/para ratio is assumed to be at thermal equilibrium according to the local gas temperature; for further details, see \citet{Woitke2009a} and \citet{Woitke2011}.

For a proper comparison to observations, we use the line radiative transfer module of \prodimo \citep[][Appendix A]{Woitke2011} to calculate the spectral line profiles for the \oiline and \htline lines at various disk \mbox{inclinations $i$}. For the local line width, we assume \mbox{$\Delta \varv=(\varv_\mathrm{therm}^2+\varv_\mathrm{turb}^2)^{1/2}$}, where $\varv_\mathrm{therm}$ is the thermal broadening and $\varv_\mathrm{turb}$ the turbulent broadening, which is fixed to $0.15\,\mathrm{kms^{-1}}$ (see also \citealt{Thi2019}). The modelled line profiles are then convolved to the spectral resolution of the observations \mbox{($R\approx50000$)}. Furthermore, we use the routine for the line profile decomposition (i.e. fitting Gaussian functions to the line profiles) of \citet{Weber2020}, which follows the fitting  approach used for the observational data \citep{Banzatti2019,Gangi2020}. Including the Gaussian fitting for the synthetic observables allows a direct comparison between the derived quantities, such as the full width at half maximum (FWHM), and the velocity peak location $\varv_\mathrm{p}$ of the Gaussian component(s) (see Sect.~\ref{sec:res_linekin}).

\begin{figure}
    \centering
    \includegraphics[width=\hsize]{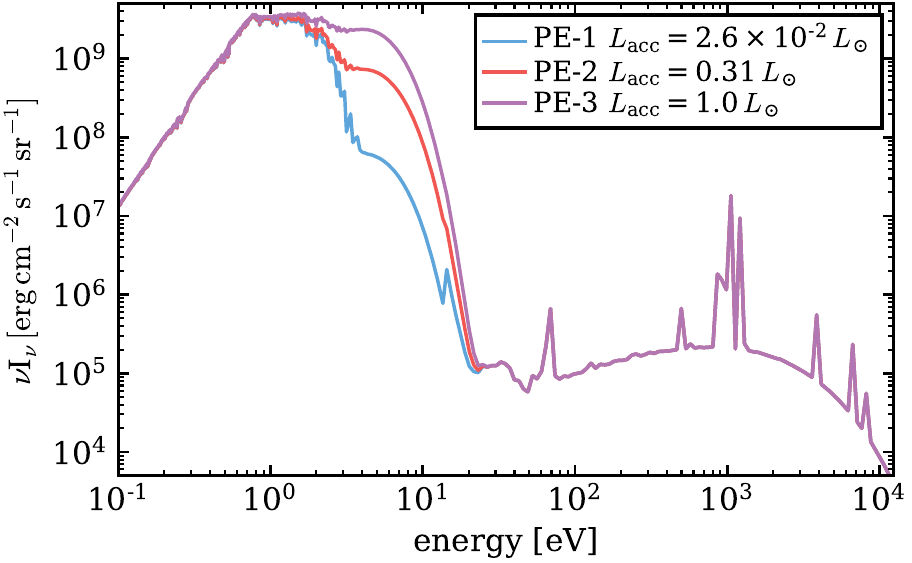}
    \caption{Stellar input spectra for the three different accretion luminosities $L_\mathrm{acc}$.}
    \label{fig:starspec}
\end{figure}
\subsection{Model series}
\label{sec:metmodelseries}
Apart from the underlying physical disk structure, we use the same stellar properties as \citet{Weber2020}. The disk is irradiated by a $M_\mathrm{*}\!=\!0.7\,M_\odot$ star using the XUV (EUV and X-rays) spectrum presented in \citet{Ercolano2008c,Ercolano2009b}. The X-ray luminosity is fixed to $L_\mathrm{X}=2\times10^{30}\,\mathrm{erg s^{-1}}$, whereas for the UV radiation field we consider three different accretion luminosities of \mbox{$L_\mathrm{acc}=2.6\times 10^{-2},\,0.31,$} and $1\,L_\mathrm{\odot}$ assuming a black-body spectrum with $T_\mathrm{eff}=12000\,\mathrm{K}$ (see Fig.~\ref{fig:starspec}). We note that we neglect any screening of the XUV emission from the star by possible accretion flows or inner MHD winds, which might be present in case of strong accretion (i.e. relevant for the PE-3 model). For screening columns $N_\mathrm{H}\gtrsim10^{22}\,\mathrm{cm^{-2}}$ , XUV photoevaporation might be completely suppressed, assuming that the screening covers all solid angles at all times \citep{Ercolano2009b}. However, this inconsistency in our models does not affect our main results significantly. In any case, and as shown below,  the high-accretion models are not expected to produce strong wind signatures for $\mathrm{H_2}$. 

\begin{table}
\caption{Model names and parameters.} 
\label{table:modelgrid} 
\centering 
\begin{tabular}{l c c c} 
\hline\hline 
Model name & $L_\mathrm{acc}$ & wind g/d ratio \tablefootmark{a} \\ 
     & [$L_\odot$] & \\ 
\hline 
PE-1 & $2.6\times10^{-2}$ & 100 \\
PE-2 & 0.31 & 100 \\
PE-3 & 1.0 & 100 \\
\hline
PE-1 gd & $2.6\times10^{-2}$ & 1000 \\
PE-2 gd & 0.31 & 1000 \\
PE-3 gd & 1.0 & 1000 \\
\hline
\end{tabular}
\tablefoot{\tablefoottext{a}{Gas-to-dust mass ratio in the wind region.}}
\end{table}
The amount of dust in the wind is relevant for $\mathrm{H_2}$ formation and for shielding of $\mathrm{H_2}$ from photo-dissociation (additionally to the self-shielding), but it is not well constrained from observations. To simulate dust entrainment in the wind, we adopt a very rough approximation and simply reduce the amount of dust in the wind region by a factor of 10 (i.e. gas-to-dust mass ratio $g/d=1000$). This is based on the dust entrainment model for photoevaporative winds of \citet{Franz2020,Franz2022} (see also \citealt{Booth2021,Hutchison2021,Rodenkirch2022}). We want to emphasise that our approach here is not entirely realistic and should only give a first indication as to the relevance of the amount of dust in the wind for the interpretation of the \htline spectral line. For this new wind dust structure, we run models again for all three accretion luminosities. The model series parameters and names are summarised in Table~\ref{table:modelgrid}. 

\begin{figure*}
    \includegraphics[width=\textwidth]{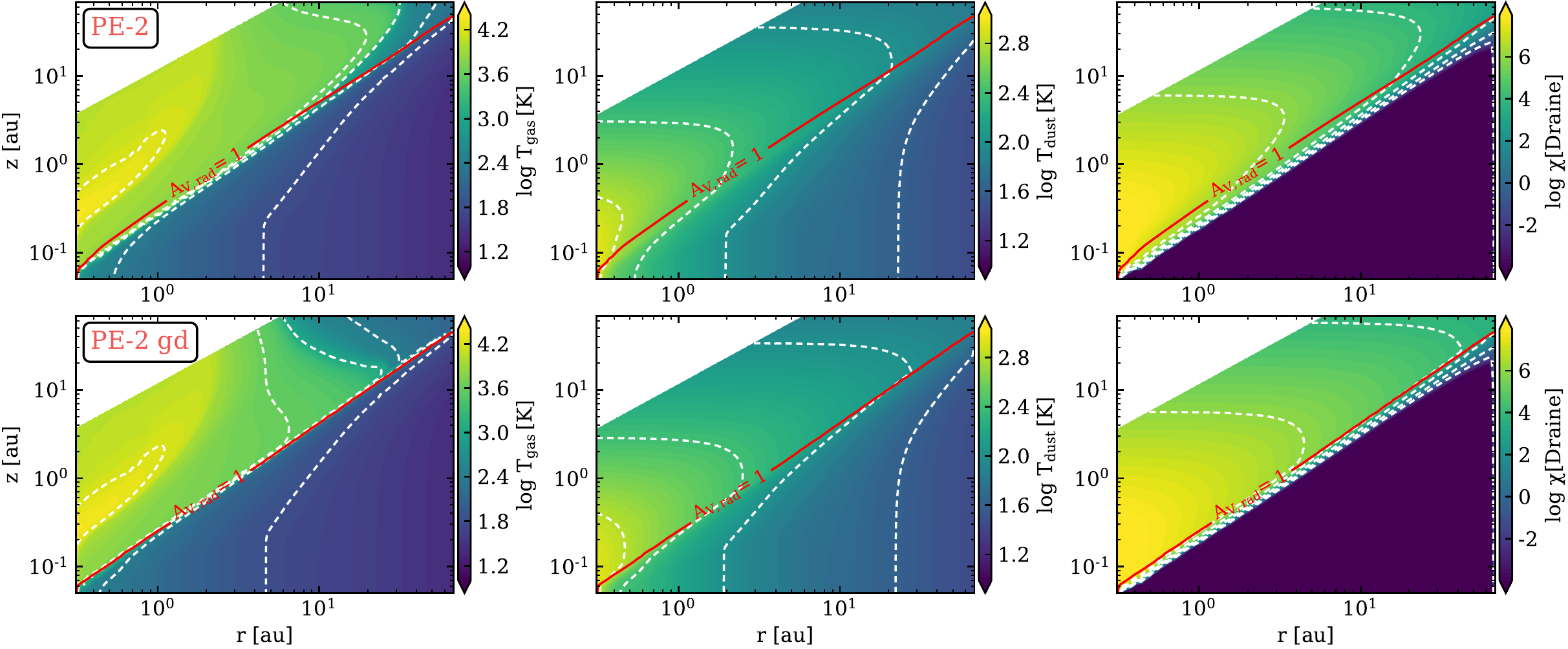}
    \caption{Two-dimensional gas temperature $T_\mathrm{gas}$, dust temperature $T_\mathrm{dust }$, and FUV radiation field $\chi$ (in units of the Draine field \citealt{Draine1996b,Woitke2009a}) for the PE-2 (top row) and \mbox{PE-2 gd} (bottom row) models. The red solid contour shows where the radial visual extinction $A_\mathrm{V,rad}$ is equal to unity, which roughly separates the wind region and the rotating disk. The white dashed contours correspond to the values shown in the colour bars.}
    \label{fig:struc}
\end{figure*}

As an example, we show physical quantities in Fig.~\ref{fig:struc} as calculated by \prodimo for the \mbox{PE-2} and the \mbox{PE-2 gd} models. Shown are the gas temperature $T_\mathrm{gas}$, dust temperature $T_\mathrm{dust}$, and the FUV radiation field $\chi$ (in units of the Draine field \citealt{Draine1996b,Woitke2009a}), which is most relevant for $\mathrm{H_2}$ photo-dissociation.
\section{Results}
\label{sec:results}
\subsection{$\mathrm{H_2}$ in the wind and the line-emitting regions}
\label{sec:lineorig}
\begin{figure*}
    \includegraphics[width=\textwidth]{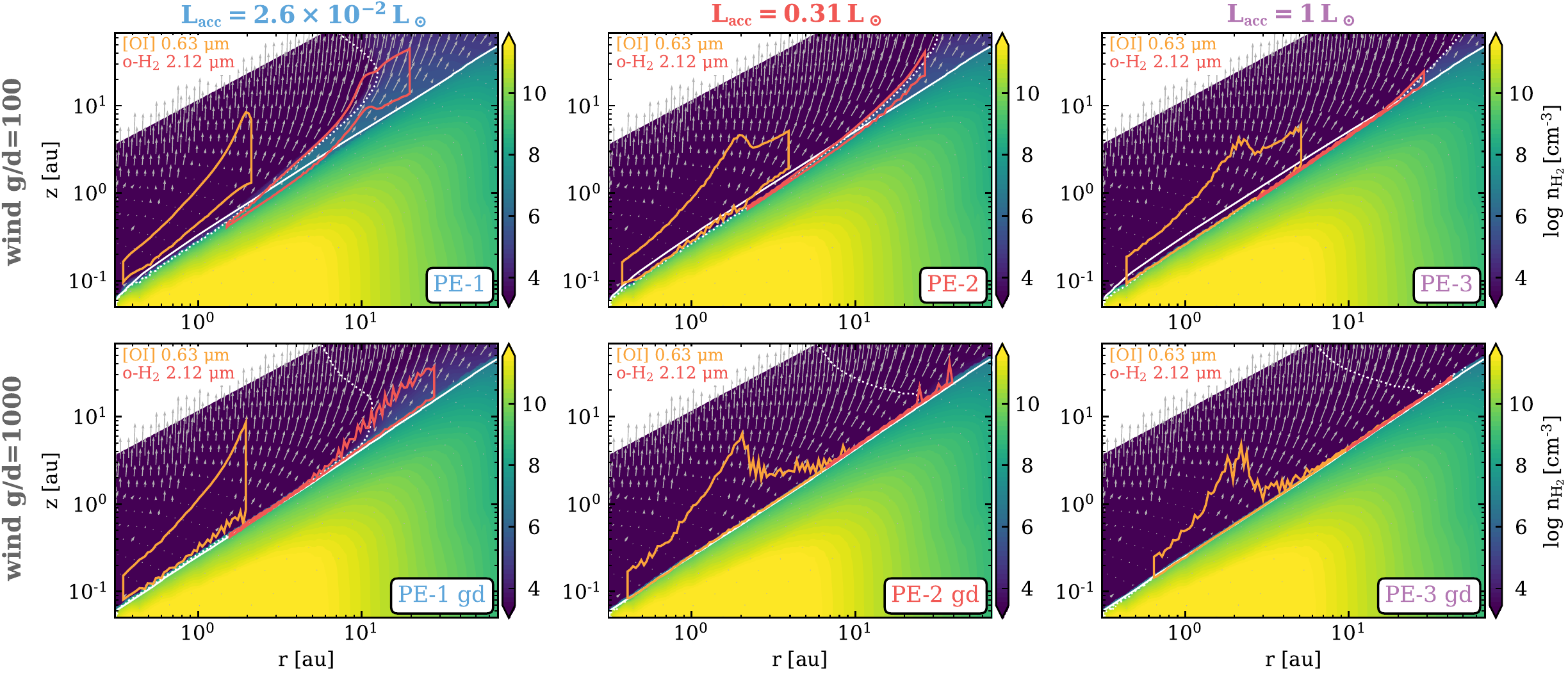}
    \caption{Main emitting region for the \oiline (orange boxes) and \htline (red boxes) spectral lines. The columns show the three models with varying $L_\mathrm{acc}$ (increasing from left to right). The top row shows the models with $g/d=100$ and the bottom row with $g/d=1000$ in the wind. The coloured contours show the molecular hydrogen number density $n_\mathrm{H_2}$. The borders of the line emission boxes indicate where the total emission reaches 15\% and 75\% in the radial (integrated inside-out) and vertical (integrated from top to bottom) directions. The solid white line shows where the radial visual extinction $A_\mathrm{V,rad}$ reaches unity, and the white dotted line $T_\mathrm{gas}=1000\,\mathrm{K}$. The grey arrows indicate the 2D velocity field (vertical and radial components of the wind).}
\label{fig:lineorig}
\end{figure*}

Figure~\ref{fig:lineorig}  shows the main line-emitting regions for \oiline and \htline for a face-on inclination. Appendix \ref{sec:physprops} (Table~\ref{table:lineorig}) also summarises the average physical properties within those line-emitting regions. 

Figure~\ref{fig:lineorig} shows that the \htline line emits from close to the disk surface (the disk--wind interface), but, depending on the accretion luminosity, a significant contribution of the line flux can  also come directly from the wind region. This implies that in those models a significant amount of molecular hydrogen can survive in the photoevaporative wind region. We emphasise that our model represents a lower limit to the $\mathrm{H_2}$ abundance in the wind as it neglects dynamical effects, and hence no $\mathrm{H_2}$ is transported from the disk into the wind. Although $\mathrm{H_2}$ would be quickly dissociated and might not survive for a long time (i.e. the photo-dissociation timescale is shorter than the flow timescale), the dynamics can only enhance the amount of $\mathrm{H_2}$ in the wind (see \citealt{Panoglou2012}) and therefore also the contribution to the total line flux from the wind region.    

For the models with less dust in the wind, the situation is similar. However, due to the lower amount of dust, FUV radiation can penetrate deeper into the wind and disk and therefore photo-dissociation of $\mathrm{H_2}$ becomes more efficient. Consequently, the main emitting region of \htline moves towards the disk surface and emission from the wind region becomes less significant. However, for the \mbox{PE-1 gd} model (lowest accretion luminosity), significant emission is still coming from the wind regions (Fig.~\ref{fig:lineorig}, bottom left panel).

Compared to \oiline, \htline is emitted from deeper layers in the disk and from farther out (up to $r\approx30\,\mathrm{au}$) in the disk. This is because $\mathrm{H_2}$ can only survive in regions where it is efficiently shielded from photo-dissociation. Although there can be some overlap in the line-emitting regions of the two lines in the radial direction, the two lines rather trace distinct regions of the disk. Furthermore, the atomic oxygen line is excited at higher temperatures (several thousand K; see Table~\ref{table:lineorig} and \citealt{Weber2020}) than \htline, which mainly comes from regions with $T_\mathrm{gas}\lesssim 1000\,\mathrm{K}$. Whether or not those distinct emitting regions can be traced with the spatially unresolved observations is discussed in Sects.~\ref{sec:res_linekin} and~\ref{sec:discussion}.
\subsection{Line luminosities}
\label{sec:linelum}
\begin{figure}
    \resizebox{\hsize}{!}{\includegraphics{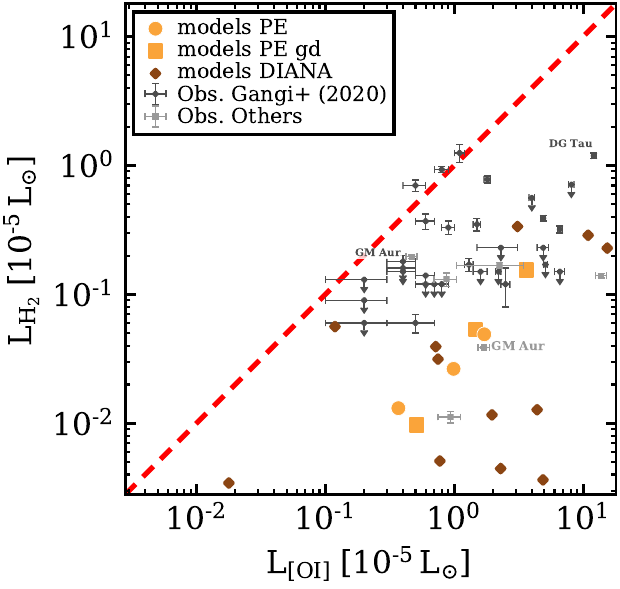}}
    \caption{\htline versus \oiline line luminosities. The dark grey points with error bars show the observations from \cite{Gangi2020,Gangi2021C}, and the light-grey points show observational data collected from the literature (see Appendix~\ref{sec:litlum} for details). The orange symbols show the photoevaporative disk-wind models (inclination $i=40^\circ$) with varying accretion luminosities (lowest(highest) line luminosities are for the lowest(highest) accretion luminosity). The square symbols are for the models with reduced dust-to-gas mass ratio in the wind. The brown diamonds are the results from the DIANA project (\prodimo models without a wind component). We mark DG Tau as an example of a target that likely cannot be modelled with a pure photoevaporative wind. We also mark the two independent observations for \mbox{GM Aur} (see Appendix~\ref{sec:litlum}).}
\label{fig:linelum}
\end{figure}
In Fig.~\ref{fig:linelum} we compare the modelled line luminosities for \oiline and \htline to the observations. We mainly use the observational data from \citet{Gangi2020}, but also discuss further data from the literature for targets where line flux measurements for both lines are available (see Appendix~\ref{sec:litlum}).

In addition to the wind models presented here, we include modelling results from the DIANA (DIsc ANAlysis) project. For this project, multi-wavelength observational data (continuum and line fluxes) for several individual disks were modelled using \prodimo and the results are published in \citet{Woitke2019}. Those disk models do not include any wind component, but can still be used to decipher the approximate impact of stellar properties and disk structure on the line fluxes. The \oiline and \htline were not included in the fitting procedure at that time, but predictions for the line fluxes were included in the models and those published values are directly used here.

From \mbox{Fig.~\ref{fig:linelum}} we see that the models populate the \htline versus \oiline line luminosity space reasonably well, but also that both the wind and the DIANA models tend to underpredict the \htline line luminosities. We discuss this in more detail in Sect.~\ref{sec:dish2lum}. For \oiline, all models are within the range of the observations, except for one outlier in the \mbox{DIANA} models. This is simply because this object, ET~Cha, is a very small ($r<10\,\mathrm{au}$) disk \citep{Woitke2011,Ginski2020}. 

The agreement of the DIANA models with the observations for three out of four targets for which observational data are available is within a factor of two to three for both lines; only for one target is the \htline  luminosity significantly lower (by
about two orders of magnitude). Overall, this is an encouraging outcome considering that the models were not optimised to fit the \htline and \oiline line luminosities. The three DIANA models with the highest \oiline and \htline luminosities are Herbig Ae/Be stars with $L_\mathrm{*}>10\,L_\mathrm{\odot}$ and these therefore also have a high luminosity in the FUV range. Observational data are not available for these targets. The \citet{Gangi2020} sample includes only one Herbig Ae/Be star, MWC 480, for which a DIANA model also exits. For MWC 480, the DIANA model also agrees within a factor of three with the \oiline luminosity, but for \htline only an observational upper limit exists, which is consistent with the model prediction. 

There is also a tendency towards slightly too low \htline luminosities for the photoevaporative wind models with respect to the observational data. However, at least for the high-accretion luminosity and/or the reduced dust-to-gas mass ratios in the wind, the predicted \htline luminosities are similar to the line luminosities observed by \citet{Gangi2020}. If the additional literature values are also taken into account, the wind models populate the \htline versus \oiline line luminosities remarkably well. We note that unlike the DIANA models, the disk-wind models use only one fiducial disk--wind structure and only the accretion luminosity is varied. The disk structure, but also the stellar properties (e.g. higher $L_\mathrm{acc}$ leads to higher line luminosities due to additional heating), can have a significant impact on the line luminosities, which is indicated by the significant scatter of the DIANA models seen in Fig.~\ref{fig:linelum}.  

Our results are also consistent with the thermo-chemical disk models of \citet{Nomura2005b} and \citet{Nomura2007ah}. \citet{Nomura2007ah} studied the impact of X-ray and FUV radiation and dust size distribution on $\mathrm{H_2}$ line fluxes and find line luminosities in the range of \mbox{$0.001\:$--$\:0.2\times 10^{-5}\,L_\mathrm{\odot}$} for \htline. This is consistent with our results, which show a range of \mbox{$0.003\:$--$\:0.3\times10^{-5}\,L_\mathrm{\odot}$} (including both wind and DIANA models; see Fig.~\ref{fig:linelum}). We note that \citet{Nomura2007ah} also studied the impact of $\mathrm{H_2}$ pumping by X-ray radiation, which is not included in our model. These authors find that as long as FUV radiation is present, X-ray pumping is not a dominating factor. However, they argue that for disks around stars with strong \mbox{X-rays} and weak UV, it might be possible to observe $\mathrm{H_2}$ emission from cooler regions excited by X-ray pumping, but more detailed modelling of this process is required. 

The comparison of the models to the data shows that the former can reproduce reasonably well the observed line luminosities for both \htline and \oiline. However, we also note that our fiducial photoevaporative disk--wind model is not a good representation of certain targets. One example is \mbox{DG Tau,} as already shown by \citet{Weber2020} for \oiline (see also \citealt{Cabrit1999}). \mbox{DG Tau} shows \oiline blueshifted emission (with respect to the systemic velocity) at velocities \mbox{$<\!-30\,\mathrm{kms^{-1}}$}, which cannot be produced by the photoevaporative disk--wind models presented here. Additionally, \htline shows strongly blueshifted emission at velocities of $10-20\,\mathrm{kms^{-1}}$ \citep{Gangi2020} and DG~Tau shows one of the highest \htline luminosities (see Fig.~\ref{fig:linelum}). Also, other observational data show indications of both MHD-driven and photoevaporative flows, although shocks might also play a role \citep[e.g.][]{AgraAmboage2014,Guedel2018}. Furthermore, \mbox{DG Tau} is strongly variable in the near- and mid-infrared \citep[e.g.][]{Varga2017,Gangi2020}. This complex environment might be the reason for the strong \htline luminosities, which cannot be matched by our models or by the models of \citet{Nomura2007ah}. Such a scenario might also be responsible for high \htline line fluxes observed in other targets.

Nevertheless, both types of models, the DIANA and the disk--wind models, predict \oiline and \htline line luminosities within the observed range. This indicates that the physical (i.e. heating and cooling), chemical, and line excitation model used in \prodimo is well suited to modelling the \htline and \oiline spectral lines, including the line kinematics.
\begin{figure*}
    \includegraphics[width=\textwidth]{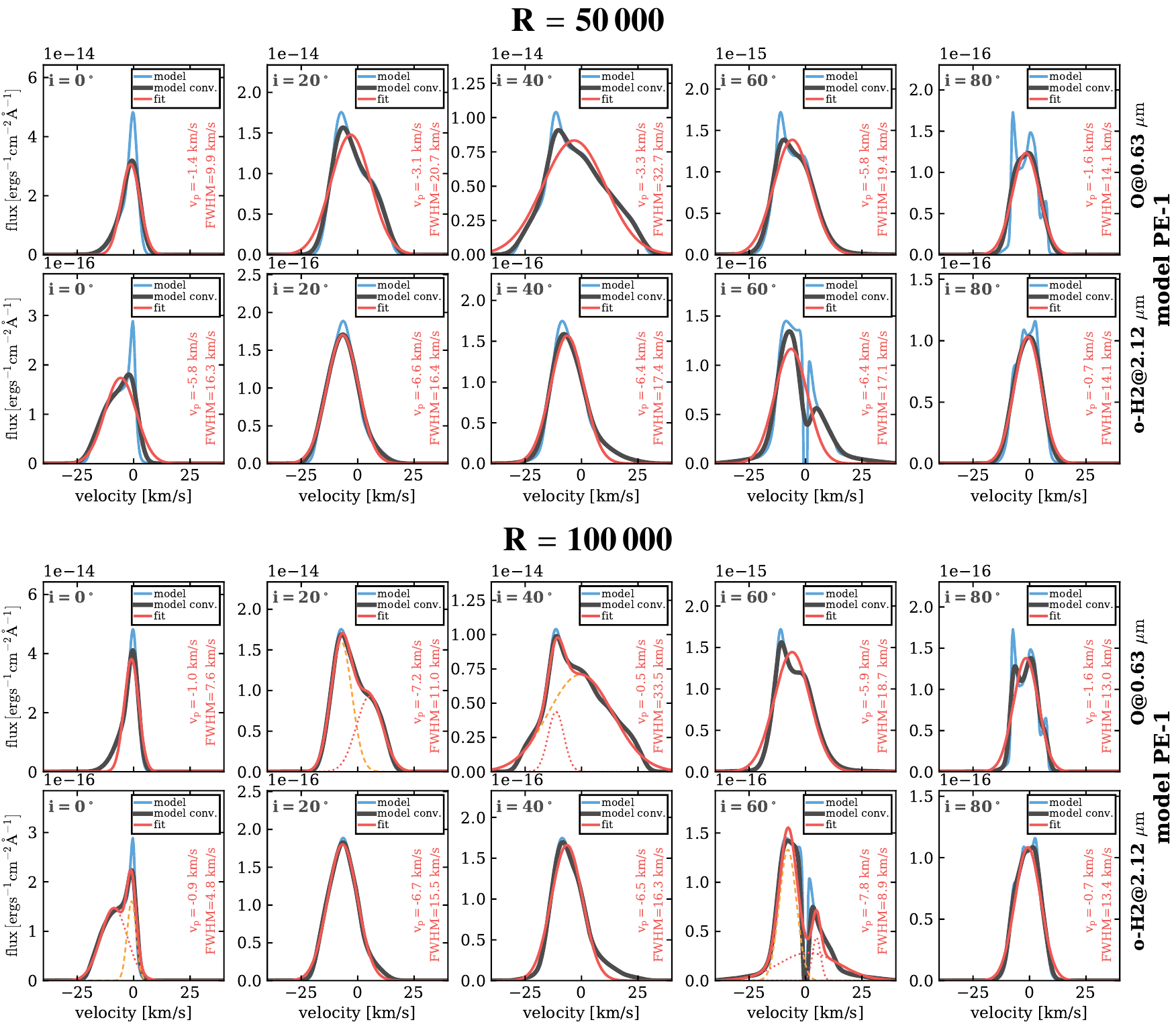}
    \caption{Example of the line-fitting results for the PE-1 model. The top figure shows the results for a spectral resolution $R=50000$ (i.e. similar to the observations). The bottom figure shows the same model, but the line profiles were convolved to a spectral resolution of $R=100000$ (see Sections \ref{sec:discusspeakvel} and \ref{sec:specrlinewidth}). 
Each individual panel shows the line profile with the model resolution (blue), convolved to the target spectral resolution (black) and the Gaussian fit (the NLVC) to the convolved profile (red solid line). In case of multiple Gaussians (dashed and dotted lines), the orange dashed line indicates the one chosen as the NLVC. This component is used to determine $\varv_\mathrm{p}$ and the FWHM (also given in each panel). The fitting results for all other models are shown in Appendix~\ref{sec:linefit}.}
\label{fig:examplefit}
\end{figure*}
\subsection{Line kinematics}
\label{sec:res_linekin}
In this section, we focus on the comparison of the kinematic properties derived directly from the observed line profiles to the model results. Those kinematic quantities are the shift of the peak location $\varv_\mathrm{p}$ and the FWHM or half-width at half-maximum (HWHM),  as presented in \citet{Gangi2020}. To derive $\varv_\mathrm{p}$ and the FWHM for the modelled line profiles, we follow the approach of \citet{Gangi2020} and fit the modelled line profiles with Gaussian components (see Sect.~\ref{sec:metlines}), focusing on the blueshifted NLVC ($\lvert \varv_\mathrm{p}\rvert < 30\,\mathrm{km s^{-1}}$). Figure~\ref{fig:examplefit} shows the results of this fitting procedure for the PE-1 model as an example. Shown are the results for two spectral resolutions $R=50000$ and $R=100000$. The first one is similar to the observations and those results for $\varv_\mathrm{p}$ and the FWHM are used for the comparison of the models to the observations. The high-resolution results are used to investigate the impact of the limited spectral resolution of the observations on the derived line kinematics and are discussed in Sections \ref{sec:discusspeakvel} and \ref{sec:specrlinewidth}, but we show them here for ease of comparison. 

As already shown in \citet{Weber2020}, photoevaporative wind models are unlikely to be the origin of the observed HVCs seen for \oiline. For \htline, \citet{Gangi2020} only considered one component, the NLVC, as none of the targets show a HVC or a clear BLVC. For comparison with the models, we only use targets from the observational sample that have a detection and derived kinematic quantities for the NLVC for both lines. From those 15 targets, 8 show $\varv_\mathrm{p}\!>\! 0\,\mathrm{kms^{-1}}$ and only 7 have a blueshifted NLVC in both lines. For the \oiline line profiles of these latter 7 targets,  two cases show a clear HVC, 2 show a BLVC, and 1 case shows no other component. For the remaining 2 cases, the high-velocity part of the profiles could not be fitted by a Gaussian (for details see Appendix~\ref{sec:targetscomp}).

As seen in Fig.~\ref{fig:examplefit} (see also Appendix~\ref{sec:linefit}), we find blueshifted components in both lines for almost all of our modelled profiles. However, especially for \htline, the derived $\varv_\mathrm{p}$ is often close to zero (see also Sect.~\ref{sec:peakvel}). Figure~\ref{fig:examplefit} also shows that at intermediate inclinations there are several cases where the profiles are not well fitted by a Gaussian (or multiple Gaussians). The reason is that, at those inclinations, the shape of the line profiles is strongly affected by the complex wind velocity field \citep[see also][]{Weber2020}, the absorption of the redshifted emission by the dust, and, in the case of \htline, also by some self-absorption. We note that this self-absorption effect is limited to  a narrow range of inclinations of about $i\approx60^{\circ}\pm5^{\circ}$ in all our models. However, the inclination at which this self-absorption effect might be seen depends on the actual disk and wind structure. The issue of non-Gaussian components in the line profiles becomes especially apparent in the models with high-spectral resolution (see Fig.~\ref{fig:examplefit}). This is discussed further in Sections \ref{sec:discusspeakvel} and \ref{sec:specrlinewidth}.
\begin{figure}
\resizebox{\hsize}{!}{\includegraphics{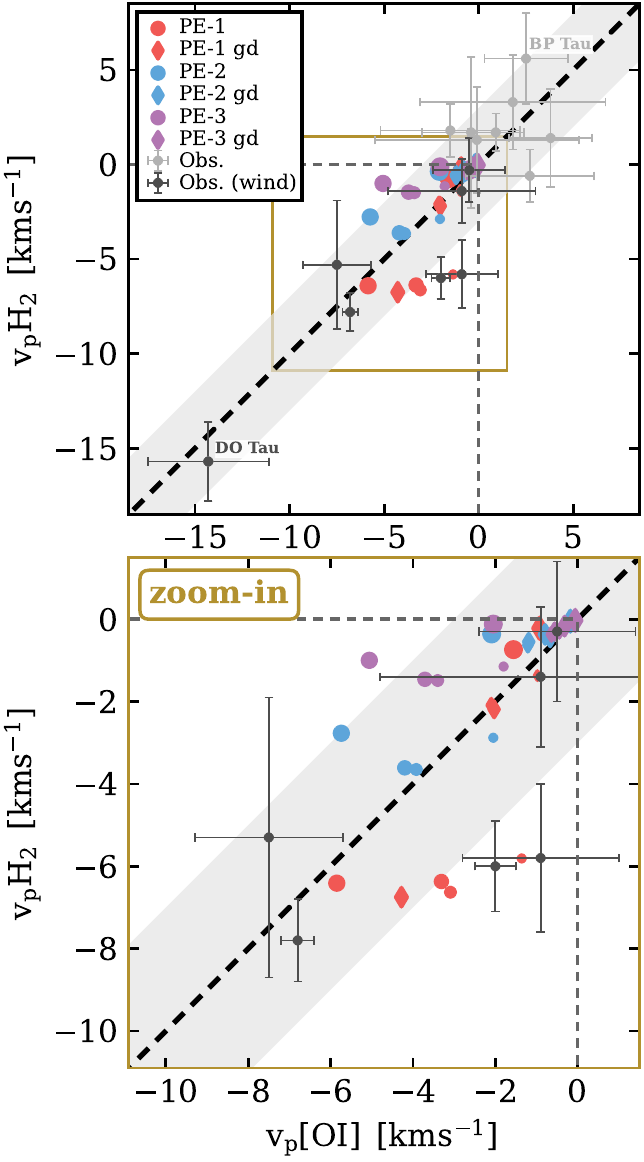}}
\caption{Velocity peak location $\varv_\mathrm{p}$ of \htline vs  $\varv_\mathrm{p}$ \oiline. Filled circles with error bars show the observed values; the black colour is for targets with $\varv_\mathrm{p}\!<\!0\,\mathrm{kms^{-1}}$ (i.e. blue-shifted wind component) for both the \htline and \oiline; light-grey marks targets with $\varv_\mathrm{p}\!>\! 0\,\mathrm{kms^{-1}}$ (i.e. no clear indication of a disk wind). The coloured symbols are for the models. Shown are models for the three different accretion luminosities and with a reduced dust-to-gas mass ratio in the wind (coloured diamonds). The size of the coloured symbols scales with the inclination; smallest symbols are for $i=0^\circ$ largest are for $i=80^\circ$. The thick dashed line indicates $\varv_\mathrm{p}\mathrm{H_2}=\varv_\mathrm{p}\mathrm{[OI]}$ and the light grey stripe indicates $\varv_\mathrm{p}\mathrm{H_2}=\varv_\mathrm{p}\mathrm{[OI]} \pm 3\,\mathrm{kms^{-1}}$. \textit{Top panel:} all data points; \emph{bottom panel:} zoom-in (brown box) showing all the models and the observational data for targets with clear wind signatures, excluding DO~Tau.}
\label{fig:vp_corr}
\end{figure}
\begin{figure}
\resizebox{\hsize}{!}{\includegraphics{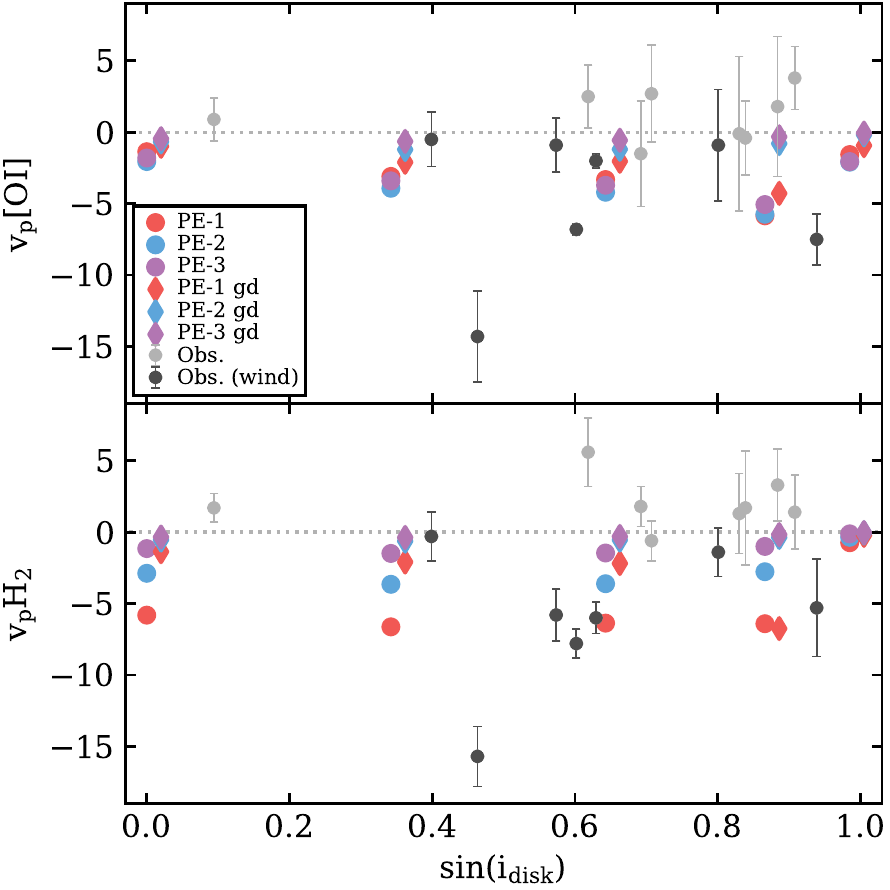}}
\caption{Velocity peak location $\varv_\mathrm{p}$ versus disk inclination for the \oiline (top panel) and \htline (bottom panel). Filled circles with error bars show the observed values; the black colour is for targets with $\varv_\mathrm{p}\!<\!0\,\mathrm{km s^{-1}}$ (i.e. blueshifted wind component) for both the \htline and \oiline; light-grey marks targets with \mbox{$\varv_\mathrm{p}\!>\!0\,\mathrm{km s^{-1}}$} (i.e. no clear indication of a disk wind). The coloured symbols are for the models. Shown are the same models as in Fig.~\ref{fig:vp_corr}.}
\label{fig:vp_inc}
\end{figure}
\subsubsection{Peak velocity}
\label{sec:peakvel}
In Fig.~\ref{fig:vp_corr} we compare the model results for the peak velocity $\varv_\mathrm{p}$ to the observations. The figure shows that the models are largely consistent with the data. One clear exception is the target DO~Tau, which shows $\varv_\mathrm{p}\!<\!-10\,\mathrm{kms^{-1}}$ for both spectral lines, whereas the largest shifts predicted in the models are  \mbox{$\varv_\mathrm{p}\!\approx\!-6\,\mathrm{kms^{-1}}$}. This is expected, as photoevaporative disk--wind models do not predict such high velocities for \oiline \citep[e.g.][]{Weber2020}. This implies that photo-evaporation is not the main physical mechanism driving the wind or outflow in DO~Tau. However, DO Tau is likely a special case as there is indication for a recent stellar encounter \citep{Winter2018} and recent observations clearly show a complex environment \citep{Huang2022}. 

Another discrepancy is that all models show $\varv_\mathrm{p}\lesssim0\,\mathrm{kms^{-1}}$ for both spectral lines, whereas for the majority of the observed targets $\varv_\mathrm{p}>0\,\mathrm{kms^{-1}}$. However, this does not necessarily mean that there is no observed signature of a wind in those targets. In particular, the observed line profiles for \oiline still show blueshifted components at high velocities, which are likely signatures of winds or outflows. However, for \htline, it is often not clear whether or not there is a blueshifted component at all, which can rather indicate that there is no significant amount of $\mathrm{H_2}$ in the wind or that the data quality is not sufficient to clearly identify the NLVC; the latter is likely also true for \oiline. In the pure photoevaporative disk--wind models it is likely easier to identify the NLVC for the \oiline compared to the observations, as the models do not include a high-velocity component. For \htline, we suspect that limitations of the data are the main cause for the group of targets with $\varv_\mathrm{p}>0\,\mathrm{kms^{-1}}$ as inspection of the individual targets does not clearly reveal redshifted emission. For example, the \htline line profile for BP~Tau, which is the target with the highest $\varv_p$, shows a high level of noise. For most of the other targets with observed $\varv_\mathrm{p}>0\,\mathrm{kms^{-1}}$ the error bars indicate that the observations are also consistent with $\varv_\mathrm{p}\approx0\,\mathrm{kms^{-1}}$, that is, no observed wind component or a slow wind. Considering those arguments, in what follows, we focus  on the targets with $\varv_\mathrm{p}\lesssim0\,\mathrm{kms^{-1}}$ for our comparisons to the observational data.

\begin{figure*}
\resizebox{0.495\hsize}{!}{\includegraphics{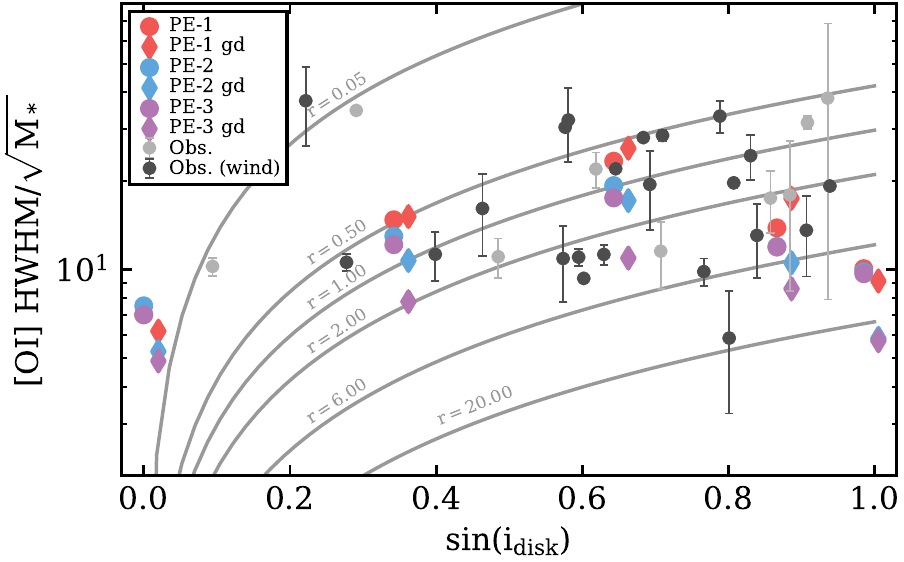}}
\resizebox{0.495\hsize}{!}{\includegraphics{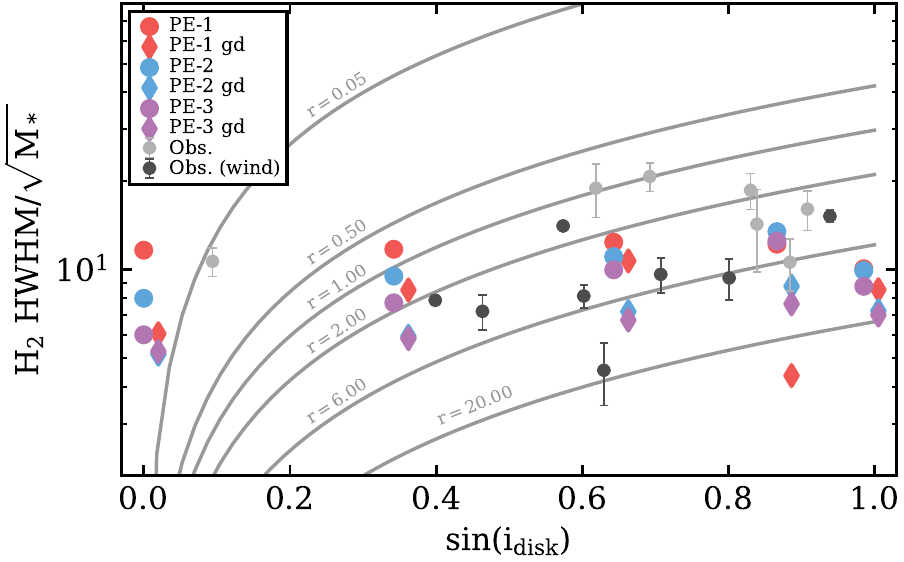}}
\caption{HWHM (left panel: [OI], right panel H$_2$) scaled by the square root of the stellar mass as a function of inclination. The coloured symbols show the models; the gd models (diamonds) are slightly shifted along the inclination axis for clarity. The black (grey) symbols show the observations, where the black coloured symbols mark the targets that show blueshifted peaks in the respective line profile. The solid lines show the maximum emission radius $R_\mathrm{K}$ derived for the given HWHM and inclination assuming Keplerian rotation for a thin disk (see Eq.~\ref{eg:RK}).}
\label{fig:HWHM}
\end{figure*}
For the models, 7 out of 30 show a \mbox{$\varv_\mathrm{p}\!<\!-3\,\mathrm{kms^{-1}}$} for \htline (we assume $\sigma (\varv_\mathrm{p})\!\approx\!3\,\mathrm{km s^{-1}}$ as a typical error in the observations). As seen from Fig.~\ref{fig:vp_corr}, this is the case for models with lower $L_\mathrm{acc}$  in particular (i.e. the PE-1 model). This is because photo-dissociation of $\mathrm{H_2}$ is less efficient because of the lower flux of FUV photons. Also, all those models have an inclination of $i\lesssim60^{\circ}$, indicating that the detection of disk winds is difficult for highly inclined disks, at least with spatially unresolved observations.  

As reported in \citealt{Gangi2020}, and as clearly seen in Fig.~\ref{fig:vp_corr}, the observational data show that \mbox{$\varv_\mathrm{p}(\mathrm{H_2})\approx \varv_\mathrm{p}(\mathrm{[OI]})$}. This is also true for the models, although there might be a slight trend towards $\varv_\mathrm{p}(\mathrm{[OI]})<\varv_\mathrm{p}(\mathrm{H_2})$ for the models with higher accretion luminosities (i.e. PE-2 and PE-3). In the framework of photoevaporative winds, such a trend would indicate that \oiline traces higher wind velocities than \htline, and that \htline is emitted from smaller heights (i.e. closer to the disk surface) than \oiline, as the wind velocity increases with height. This is indeed the case for our photoevaporative wind models, as seen in Fig.~\ref{fig:lineorig} and discussed in Sect.~\ref{sec:lineorig}.

According to \citet{Gangi2020}, there is no correlation between $\varv_\mathrm{p}$ and disk inclination $i$. As we present only one physical wind--disk structure, the model series is not well suited to investigating this non-correlation. Nevertheless, it is still interesting to check the existing models for a correlation with $i$ . As seen in Fig.~\ref{fig:vp_inc}, there is a slight trend of increasing $\varv_\mathrm{p}$ with inclination within each group of models with the same $L_\mathrm{acc}$, that is, if the highest inclination is excluded (those models always show $\varv_\mathrm{p}\approx 0\,\mathrm{km s^{-1}}$), where the trend is stronger for \oiline than for \htline. Considering the expected errors in real observations, such a correlation would be hard to identify (see also \citealt{Gangi2020}). Furthermore, Fig.~\ref{fig:vp_inc} also shows that both $L_\mathrm{acc}$ and the gas-to-dust mass ratio have a stronger (or at least similar) impact on $\varv_\mathrm{p}$ than the inclination itself. Although we present only one physical structure, our model series indicates that it is challenging to observe a clear correlation of $\varv_\mathrm{p}$ with inclination and that other properties of the targets, such as the accretion luminosity,  also have to be considered. 
\subsubsection{Full width at half maximum}
The FWHM or HWHM of a spectral line can be used to determine the maximum radius of the emitting region of the line. \citet{Gangi2020} used the relation 
\begin{equation}
\label{eg:RK}
R_K={\left(\frac{\sin(i_\mathrm{disk})}{\mathrm{HWHM}}\right)}^2\times G\times M_*,
\end{equation}
where $G$ is the gravitational constant and $M_*$ is the stellar mass. This relation assumes a thin disk in Keplerian rotation. This approximation does not fully represent the physical conditions for the spectral lines studied here, as the lines are not emitted in the midplane or in a thin layer, and because the spectral line profiles are strongly affected by the wind velocity \citep[see][]{Weber2020}. Nevertheless, it is a rough indicator of the maximum emitting radius $R_\mathrm{K}$ and it can be  directly applied to observational data. To allow for a direct comparison between the models and the data of \citet{Gangi2020}, we also use Eq.~\ref{eg:RK} and measure the HWHM of the modelled spectral lines from the fitted Gaussian of the respective NLVC. 

In Fig.~\ref{fig:HWHM}, we compare the derived HWHM scaled by the stellar mass as a function of disk inclination to the results from \citet{Gangi2020}. The model results are in good agreement with the data for both \oiline and \htline. The measured HWHM from the modelled \oiline profiles show a peak at medium inclination ($i=40^\circ$), but the scatter is also significant, especially if the models with lower dust content in the wind are also considered. Such a trend is not really visible in the data, but the error bars are large and only very few measurements for disks close to face-on or edge-on exist. For \htline, the HWHM does not show a dependence on disk inclination either in the models or in the available observational data.

\mbox{Figure~\ref{fig:HWHM}} also indicates that, for the models, the HWHM for \oiline is systematically larger than for \htline in the range of $\sin(i)=0.2 - 0.9$. For those 16 models, the ratio $\mathrm{HWHM}_\mathrm{[OI]}/\mathrm{HWHM}_\mathrm{H_2}$ ranges from $\approx0.9$ to $\approx2.4$, except for one outlier with a ratio of four and two models with a ratio of smaller than unity. The situation is similar for the subset of the observational data with $\varv_\mathrm{p}<0\,\mathrm{km s^{-1}}$  in both lines (all $\sin(i)$ are within 0.2 and 0.9). For those targets, the ratio $\mathrm{HWHM}_\mathrm{[OI]}/\mathrm{HWHM}_\mathrm{H_2}$ is in the range of \mbox{$\approx\!0.6$} to \mbox{$\approx\!2.5$}, and 2 out of 7 have a ratio smaller than unity. We further discuss this in Sect.~\ref{sec:discussionkin}.
\section{Discussion}
\label{sec:discussion}
\subsection{Interpretation of observed line kinematics}
\label{sec:discussionkin}
As discussed in \citet{Gangi2020}, the observational data show that the line kinematics for the two spectral lines \oiline and \htline are similar. \citet{Gangi2020} argue that this might indicate that the lines trace similar physical regions of a disk wind, and that such a scenario is more consistent with a centrifugal MHD-driven wind as studied in \citet{Panoglou2012}. \citet{Panoglou2012} follow the thermo-chemical evolution (including FUV and X-ray irradiation) along wind streamlines derived from MHD models. These authors found that the region where $\mathrm{H_2}$ can exist in the wind evolves with the evolutionary stage (i.e. Class 0/I/II) of the target; in particular, wind temperatures become higher and the shielding of $\mathrm{H_2}$ less efficient because of the stronger FUV radiation field impinging on the disk at later stages. We note that \citet{Panoglou2012} did not produce synthetic observables for their models, and therefore a direct comparison to observational data is not possible. Nevertheless, their results for the Class~II/T~Tauri stage are in general agreement with our results, as we also find that molecular hydrogen only survives close to the disk surface.

In this work, we show that observed line kinematics for \oiline and the \htline are consistent with a photoevaporative wind. This indicates that both scenarios, that is, MHD- and photoevaporative driven winds, might be consistent with the kinematics derived from currently available \htline observations. The main argument from \citet{Gangi2020} for MHD winds is that the observational data seem to indicate that the \oiline and \htline are tracing similar regions of the wind. This is in contradiction to our modelling results, which indicate that \oiline comes from regions closer to the star and higher up in the wind compared to the emitting region of \htline (see Fig.~\ref{fig:lineorig}). Nevertheless, the kinematic quantities derived from our models are roughly consistent with the data. There may be several reasons for this: the spectral resolution of the observational data is not high enough to discriminate the two emitting regions; the thin-disk approximation to determine the line emitting region is perhaps too simple, and the physical properties (i.e. the wind velocity) are similar in both line-emitting regions (i.e. because of different heights). In the following sections, we discuss these possibilities in more detail, and separately for the peak velocity and FWHM.
\subsubsection{Impact of spectral resolution on peak velocity}
\label{sec:discusspeakvel}
\begin{figure}
\resizebox{\hsize}{!}{\includegraphics{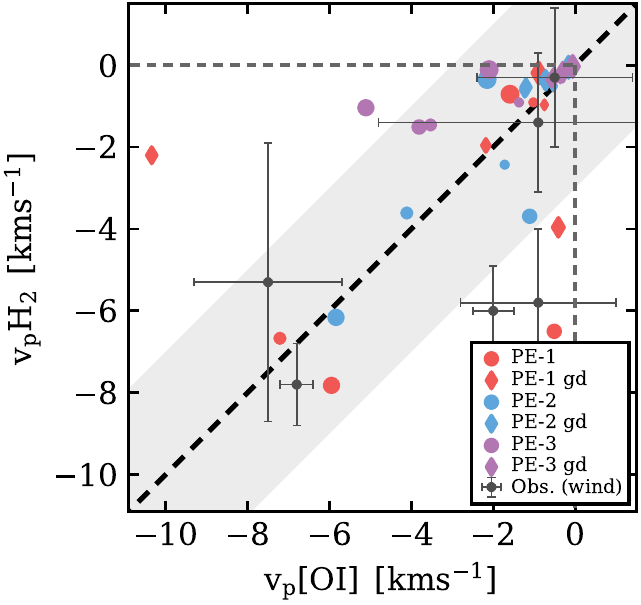}}
\caption{Same as the bottom panel of Fig.~\ref{fig:vp_corr}, but for the modelled line profiles a spectral resolution of $R=100000$ is used.}
\label{fig:vp_corrHR}
\end{figure}
To determine the shift in the peak velocity $\varv_\mathrm{p}$, the spectral resolution is most relevant. As seen from the data, many targets are around $\varv_\mathrm{p}\approx0\,\mathrm{km s^{-1}}$ with error bars as big as $\pm 3\,\mathrm{km s^{-1}}$, and therefore it is unclear whether there is a wind signature or not. 

Using the models, we have the possibility to study the impact of the spectral resolution. Figure~\ref{fig:vp_corrHR} shows the same observational data as in Fig.~\ref{fig:vp_corr} but for the models we now assume a spectral resolution of $R\!=\!100000$, which is two times higher than for the data. The general picture for the models does not change significantly. As can be seen in Fig.~\ref{fig:vp_corrHR}, some models now show $\varv_\mathrm{p}\lesssim -3\,\mathrm{kms^{-1}}$ for one line but $\varv_\mathrm{p}\approx 0\,\mathrm{kms^{-1}}$ for the other, which increases the scatter in the $\varv_\mathrm{p}\mathrm{H_2}$ versus $\varv_\mathrm{p}\mathrm{[OI]}$ plane. This is a consequence of the fitting procedure. For those models, the line profiles are now fitted by two (or more) Gaussian components, instead of one in the low-spectral-resolution case. Especially for the PE-1 model, we find that out of the ten line profiles, four are fitted by multiple Gaussians for $R\!=\!100000$ (see Fig.~\ref{fig:examplefit}). For example, for $i=0^{\circ}$, the PE-1 model shows a narrow peak and a significantly blueshifted component due to the wind (first column in Fig.~\ref{fig:examplefit}).  This profile was still fitted by a single Gaussian in the low-resolution case, but  two Gaussians were identified in the high-resolution case. This makes the identification of the NLVC in the model ambiguous. A similar issue arises at higher inclinations for \htline because of self-absorption (see Sect.~\ref{sec:res_linekin}).

We note that for the models with $R=50000,$ the identification of the NLVC is straightforward as all modelled profiles are fitted with a single Gaussian except for one case. In the \mbox{PE-1 gd} model, the line profile for the \htline at $i=60^{\circ}$ (see Fig.~\ref{fig:linefitgd}) requires a two-Gaussian fit, but this is due to the self-absorption effect. For the high-resolution models, 9 out of the 60 modelled line profiles resulted in a multi-Gaussian fit. Most of these are for the PE-1 and \mbox{PE-1 gd} models (7 out of 20), where we see the strongest wind signatures in the line profiles. Although an unambiguous identification of the NLVC
is possible for the majority of our models  by applying the Gaussian fitting procedure, our results also indicate that a clear identification of the NLVC might not be feasible for higher spectral resolution observations. For high-resolution observational data, the identification of the NLVC will likely be even more ambiguous, in particular for the cases where broad and high-velocity components are also present, as they might not be well represented by Gaussian components at all. On the other hand, such high-resolution observations might allow more detailed studies of the line shape without Gaussian fitting; for example, the blueshift of the profile could be  determined simply by measuring the velocity at the actual peak of the line profile. 

For this experiment, we did not adapt the fitting procedure, and the fitting mechanism selects the component with $\varv_\mathrm{p}$ closer to zero. This is rather an issue of the definition of the NLVC and the Gaussian-fitting procedure, which might need to be revised for higher spectral resolution data. However, it does not affect the conclusion that, in the models, both lines still show signatures of blueshifted emission. Nevertheless, this exercise shows that an increase in the spectral resolution by a factor of two does not change the general picture that, for both the models and the data, the derived $\varv_\mathrm{p}$ is similar for both lines in many cases.

\subsubsection{Interpreting the observed line width}
\begin{figure}
    \resizebox{\hsize}{!}{\includegraphics{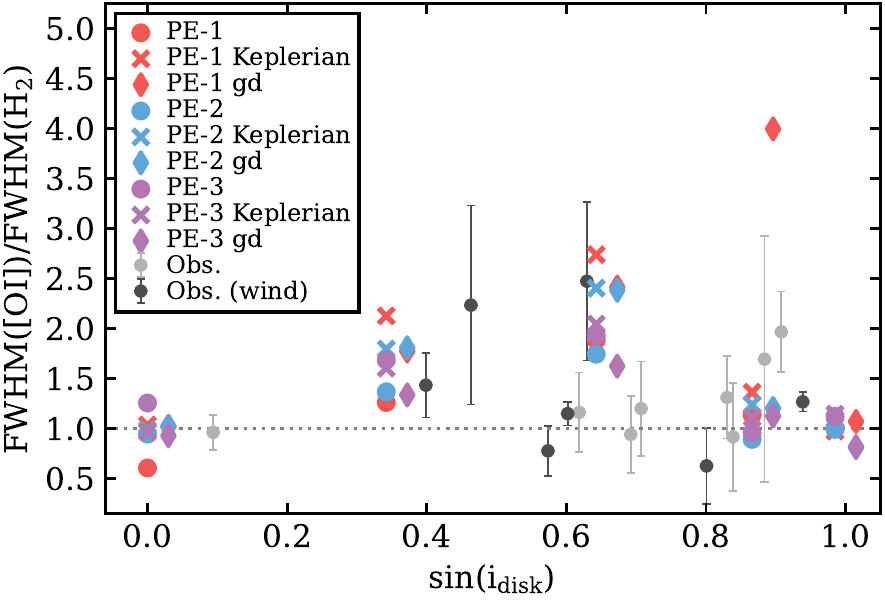}}
    \caption{Ratio of the FWHM of \oiline and \htline. Shown are the same models as used for Fig.~\ref{fig:HWHM}. Additionally, we show the model series called \textit{Keplerian} (crosses), where we assume that the velocity field in the disk and wind is purely Keplerian (i.e. thin-disk approximation). The black data points with error bars mark the targets with blueshifted $\varv_\mathrm{p}$ in both lines.} 
\label{fig:FWHM_ratio}
\end{figure}
Figure~\ref{fig:FWHM_ratio}  shows the ratio of the FWHM of the two studied lines to compare the measured line widths. The model results show that the FWHM for \oiline is larger than that for \htline for $20^{\circ}<i<60^{\circ}$. This is also the case in the data but is not as evident there. In particular, at $\sin(i)\approx0.6,$ there are two targets with a ratio close to one: \mbox{XZ Tau} and \mbox{DG Tau}. \mbox{XZ Tau} has a ramp-like \oiline line profile and therefore has some uncertainty in the fit \citep{Gangi2020}. DG Tau is likely not well represented by our model, as already discussed in Sect.~\ref{sec:linelum}. The one target with a ratio $<1$ at $\sin(i)\approx0.8$ is \mbox{GM Aur}, which is a transitional disk with a large inner hole and rings (see e.g. \citealt{Huang2020}), which might be the reason for the observed narrow and similar line widths for the \oiline and \htline (see also discussion in \citealt{Gangi2020}); however, at those high inclinations, our model also predicts ratios close to unity.

As already mentioned, a direct comparison of individual targets to our models is not feasible, as here we only present a fiducial model. Nevertheless, Fig.~\ref{fig:FWHM_ratio} clearly shows that the photoevaporative wind models are consistent with the observationally derived kinematics, even though in our models the \oiline and \htline lines are not emitted from the same region in the wind or disk. In the following, we want to focus on this aspect and use our models to get a better understanding of the limitations that the simple Keplerian thin-disk approximation and the limited spectral resolution put on our interpretation of the line width.
\subsubsection{Impact of the wind velocity on the line width}
\label{sec:diswinvel}
The presence of a wind can affect the shape of the line profile and therefore also affects the measured FWHM in the disk. This was already discussed in detail by \citet{Weber2020} for \oiline and other atomic tracers. An important conclusion of \citet{Weber2020} is that in most cases the width of the emission lines is not a good indicator of the physical region from where the lines were emitted. At low inclinations, the line width is not determined by Keplerian rotation, and at higher inclinations, different velocity components are projected into the same observed velocity bin, which makes the localisation of the physical emitting region difficult. 

For the wind tracers studied in this work, the most interesting case is the face-on view ($i=0^{\circ}$), for which Keplerian broadening does not play a role (e.g. TW~Hya). As can be seen from Fig.~\ref{fig:FWHM_ratio}, in the model PE-1 (lowest $L_\mathrm{acc}$), the measured FWHM of \htline is larger than for \oiline. This is caused by the traced wind velocities. In the PE-1 model, $\mathrm{H_2}$ survives higher up in the wind region (see Fig.~\ref{fig:lineorig}) than in the other models, and consequently the profile is dominated by the wind velocities, causing the line broadening (see top-panel of Fig.~\ref{fig:linefit}). The average wind velocity in the \htline -emitting region is $\approx 2.5$ times higher than in the region emitting \oiline (see Table~\ref{table:lineorig}). The opposite is true for the PE-3 model (highest $L_\mathrm{acc}$), where $\mathrm{H_2}$ is more efficiently dissociated and the line emission origin moves towards the disk surface where the wind velocities are approaching zero. In the models with reduced dust mass in the wind, the FWHMs of the two lines are almost identical. In those models, both line-emitting regions are closer to the disk surface and trace generally lower wind velocities, and the impact on the line broadening is limited. Nevertheless, the models predict that the ratio of the FWHM also depends on the stellar properties (at least at low inclinations) and show the potential of using atomic and molecular tracers for  studying the wind velocity structure in different regions. However, a comprehensive study of this effect requires more observations of targets with varying stellar properties and low-inclination disks.

For the two lines studied here, the impact of the wind velocities on the FWHM should be similar, and hence it is unlikely that this is why we measure a similar FWHM for the two lines using the Keplerian thin-disk approximation. 
\begin{figure}
\resizebox{\hsize}{!}{\includegraphics{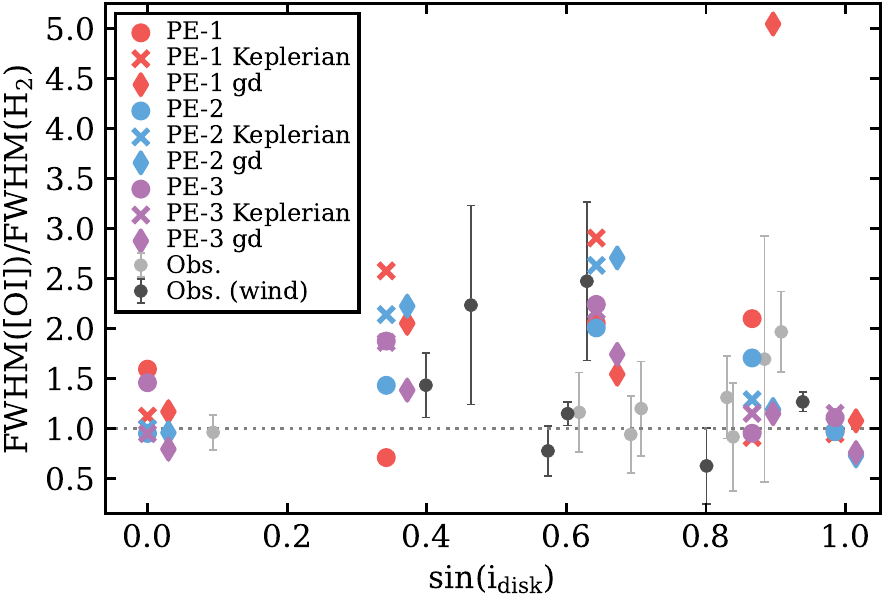}}
\caption{Same as Fig.~\ref{fig:FWHM_ratio} but a spectral resolution of R=100000 is used  for the modelled line profiles.} 
\label{fig:FWHM_ratioHR}
\end{figure}
\subsubsection{Impact of the spectral resolution on line width}
\label{sec:specrlinewidth}
The measured FWHM of the line profiles also depends on the spectral resolution of the observations. Fig.~\ref{fig:FWHM_ratioHR} shows the FWHM ratio of the two lines  again but the modelled line profiles are now convolved to a spectral resolution of $R=100000$ (i.e. the same models as discussed in Sect.~\ref{sec:discusspeakvel}; see also Fig.~\ref{fig:examplefit} for an example). A comparison with the results shown in Fig.~\ref{fig:FWHM_ratio} (low resolution) shows that, for 14 models the ratio increases, for 6 it decreases, and for 10 it remains similar (within 5\%). The biggest changes are for $\sin(i)\approx0.85$ where we now see clear differences in the FWHM of \oiline and \htline, but for the models with reduced dust content the ratios at that inclination are still similar. Nevertheless, for the modelled photoevaporative wind scenario, the lower spectral resolution of $R\approx50000$ tends to make the FWHM of the two lines more similar, and higher spectral resolution observations would be helpful in order to better constrain the line origin. However, the general picture for the models remains the same, in particular the impact of the wind velocity on the FWHM, as discussed in Sect.~\ref{sec:diswinvel}. 
\subsubsection{The thin-disk approximation}
\label{sec:disthindisk}
In addition to the wind velocity field, the Keplerian thin-disk approximation used for Eq.~(\ref{eg:RK}) also neglects the vertical position of the emission region in the disk, with the latter actually affecting the rotation velocity \citep[e.g.][]{Rosenfeld2013}. Looking at the physical emitting regions in the model (see Fig.~\ref{fig:lineorig}), one can see that \oiline \ and \htline are emitted from different heights. This means that even if the two lines emit from the same radius, the measured velocities will be different.

Fig.~\ref{fig:FWHM_ratio} also shows the FWHM ratios for \mbox{PE-2} models, where we replace the actual velocity field with pure Keplerian rotation, also neglecting the height of the disk (models with suffix Keplerian). This should simulate the thin-disk approximation, although we note that, in the model, the emission is still vertically extended. To derive the FWHM for those pure Keplerian line profiles, we followed the same Gaussian fitting procedure as for the wind models (see Fig.~\ref{fig:linefitK}).

Figure~\ref{fig:FWHM_ratio} shows that for pure Keplerian models, the FWHM ratio for the PE-3 model series remains similar to that in the wind models. The reason is that the PE-3 models show the weakest wind features in the line profiles. For both lines, the emission is coming from closer to the disk surface, and in case of \htline there is less $\mathrm{H_2}$ in the wind due to the strong FUV radiation field (see Fig.~\ref{fig:lineorig}). Therefore, the line profiles are mostly dominated by the Keplerian velocity field of the disk in any case. 

For the \mbox{PE-1} and \mbox{PE-2} models, the FWHM ratio increases for $i \approx 20^\circ - 60^\circ$ by factors of up to $\approx1.7$ and $\approx1.4$, respectively. The PE-1 models show the strongest wind features in the line profiles, especially for \htline, as more $\mathrm{H_2}$ can survive in the wind causing broader line profiles in general. In the Keplerian models, this broadening due to the wind is neglected and therefore the FWHM ratio increases.  This implies that considering a more realistic velocity field (such as that in the models) brings the measured FWHMs for the lines closer to each other and therefore makes their emitting radii appear more similar, although they are not.

Nevertheless, the general picture for the measured FWHM in the models remains similar, and is also consistent with the observational data for the FWHM. This means that the trend in the FWHM ratio of the models is mainly driven by inclination. This is a consequence of the different physical emitting regions. The \oiline always comes from radii $\lesssim 2\,\mathrm{au}$ tracing higher rotation velocities, whereas the \htline flux is dominated by the emission from a larger area and only relatively little emission is coming from the inner region. For increasing inclinations, the measured projected velocities increase and hence the line gets broader. Although this happens for both lines, for \oiline this high-velocity emission from the inner disk has a stronger impact on the spatially integrated line profile as the total emitting area remains small.

\begin{figure}
\resizebox{\hsize}{!}{\includegraphics{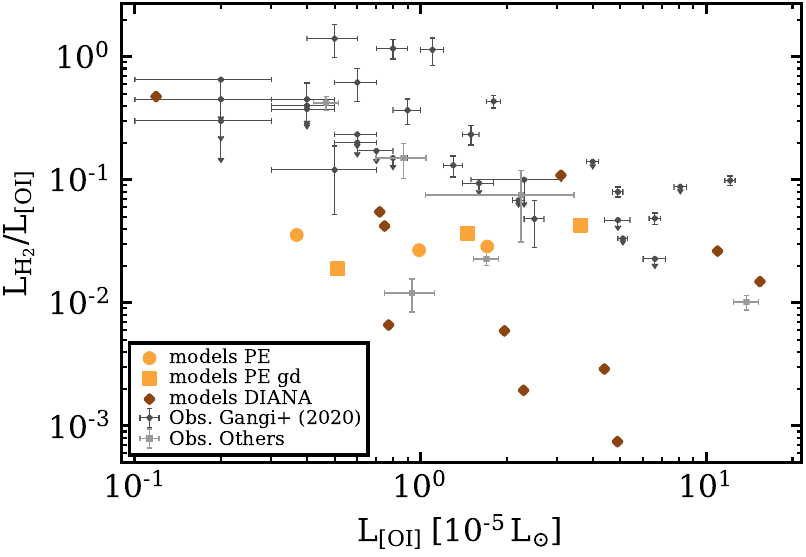}}
\caption{Line luminosity ratio of \htline and \oiline as a function of the \oiline luminosity. The observational data points and models shown are the same as in Fig.~\ref{fig:linelum}. We omit the one DIANA model with the lowest line luminosities for clarity. The upper limits for the line ratios (downward arrows) are a consequence of the \htline line luminosity upper limits.}
\label{fig:lineratios}
\end{figure}
\subsection{The \htline luminosity}
\label{sec:dish2lum}
As mentioned above, our models tend to systematically under-predict the \htline line luminosities (see Sect.~\ref{sec:linelum}). Although we do not expect this to impact our conclusions on the line kinematics, we feel it necessary to discuss the potential reasons for the overly low \htline luminosities and possibilities for improvements of the models. In Fig.~\ref{fig:lineratios}, we present the luminosity ratio of \htline to \oiline  ($L_\mathrm{H_2}$/$L_\mathrm{[OI]}$) as a function of $L_\mathrm{[OI]}$, as this more clearly shows the overly low $L_\mathrm{H_2}$ with respect to $L_\mathrm{[OI]}$ in the models. It is apparent from Fig.~\ref{fig:lineratios} that especially for $L_\mathrm{[OI]}$ around $10^{-5}\,L_\mathrm{\odot}$ our models under-predict $L_\mathrm{H_2}$  by about an order of magnitude. 

One reason might be that, for such cases, the layer  emitting the majority of  \htline is colder than the \oiline-emitting regions, pointing towards an inaccurate vertical temperature gradient in our models. Further investigation of this potential issue would require a detailed study of the heating and cooling mechanism for models with varying stellar and disk properties that cover this observational parameter space. Furthermore, the detailed line excitation mechanisms for \htline might need to be re-evaluated (see e.g. \citealt{Nomura2005b,Nomura2007ah}), but such investigations are out of the scope of this paper. 

Another possibility is that for targets with high $L_\mathrm{H_2}$/$L_\mathrm{[OI]}$, the \htline line luminosity also includes contributions from outflows and jets (e.g. DG Tau). The spatially resolved \htline observations of \citet{Beck2019} indicate a complex spatial distribution  of the \htline line emission for some targets, indicating shock-excited emission. Although such emission should usually show higher velocities, a potential contribution to the NLVC cannot be excluded. We note that, for GM Aur, the observations of \citet{Beck2019} indicate compact and centralised emission, pointing towards a disk origin. For this target, the predicted \htline flux of the GM Aur DIANA model is in reasonable agreement with the observations (within a factor of three), indicating that for such targets our model works quite well. 

As the \htline traces non-Keplerian velocity fields in several cases, dynamical effects might also be highly relevant. For the wind scenario, additional molecular hydrogen could be transported into the wind regions and might survive long enough to contribute significantly to the line flux. However, even without a disk wind,  dynamical effects such as advection of gas through the ionisation front at the $\mathrm{H^+}/\mathrm{H}/\mathrm{H_2}$ transition, as studied by \citet{Maillard2021} for photo-dissociation regions, might also have significant effects for the disk scenario. However, to study such effects, more sophisticated models are required (see discussion in Sect.~\ref{sec:disfuture}). 

Another important point is that both the observational sample and the presented models are biased. From Fig.~\ref{fig:linelum}, it is clear that the observations of \citet{Gangi2020} are only sensitive down to $L_\mathrm{H_2}$ of $\approx10^{-6}\,L_\mathrm{\odot}$ and therefore might miss a large sample of disks with low $L_\mathrm{H_2}$ (i.e. such as TW Hya). The bias in the model stems from the facts that we only present a fiducial disk--wind model and the DIANA models were not selected to cover the physical properties of the \citet{Gangi2020} sample. To better understand the origin of the \htline emission and the shortcomings of the models presented here, both increasingly sensitive observations and a detailed modelling of selected targets covering the observed range of physical properties are necessary.
\subsection{Model limitations and future aspects}
\label{sec:disfuture}
In this work we focus on observables of disk winds from molecular and atomic species by post-processing radiation--hydrodynamical simulations with a thermochemical code. Although the radiation--hydrodynamic models used consider thermochemical processes for atoms (i.e. photo-ionisation), molecules are neglected and are only included in the post-processing step. However, thermochemical processes can also influence the physical disk-wind properties; for example molecular cooling might affect the thermal structure and hence the wind-launching region. 
There are a few models that include thermochemical processes in the (magneto) hydrodynamic modelling, compensating for the high computational cost with drastic approximations in the radiative transfer  \citep[e.g.][]{Wang2017a,Nakatani2018a,Nakatani2018,Wang2019,Gressel2020}. The majority of these models are not suitable for producing synthetic observations, given the uncertainties on the derived temperature structures, and this was only attempted in \citet{Gressel2020}, although without a direct comparison to existing observations. 

While molecular cooling is not expected to affect the wind structure and mass-loss rates significantly \citep{Owen2010,Sellek2022}, a computationally efficient self-consistent approach with an appropriate treatment of the radiative transfer would be highly desirable. To that end, we are in the process of coupling the existing hydrodynamic photoevaporative wind models with the thermochemical code PRIZMO \citep{Grassi2020}, especially designed for this purpose. We will make use of the existing models such as \mbox{MOCASSIN} and \prodimo to verify the new code but also to identify the most relevant thermochemical processes that need to be considered in a self-consistent model, the aim being to build a computationally efficient model that allows for direct comparison to observations. 

Furthermore, modelling of various potential wind tracers is required. To the best of our knowledge, this study is the first attempt at comparing disk-wind model predictions for both atomic and molecular wind tracers directly to observations. However, to fully understand disk winds and their origin, multi-wavelength studies and observations are necessary. For example, the models of \citet{Gressel2020} indicate that the CI line at 492 GHz ($609\,\mathrm{\mu m}$) is a tracer that could  potentially be used to tell apart a magnetocentrifugal from a photoevaporative wind. More recently, \citet{Xu2021} presented observations of the \mbox{CII} resonance line at $1335$\AA\  of a sample of protoplanetary disks. These authors interpret the observed absorption line profiles with a semi-analytic thermal-magnetic wind model, indicating that both wind mechanisms might be important. Furthermore, CO ro-vibrational lines in the mid-infrared also show signatures of winds and outflows (\citealt{Banzatti2022} and references therein). However, modelling these various disk wind tracers with self-consistent models could be challenging or even unfeasible due to computational limitations. 

A similar approach as used in this work might be preferable, and we plan to use existing dynamic wind-model grids \citep[e.g.][]{Picogna2021} to explore the impact of disk structure (e.g. transitional disks) and stellar properties on atomic and molecular observables in various wavelength regimes. This will also include predictions for spectrally and spatially resolved observations (including spectro-astrometry, e.g. \citealt{Whelan2021}) of disk wind tracers. Currently, this is only feasible for molecular hydrogen to some extent (see \citealt{Beck2019}), but upcoming facilities such as the Extremely Large Telescope (ELT) and potentially \mbox{GRAVITY+}\footnote{\url{https://www.mpe.mpg.de/ir/gravityplus}} at the Very Large Telescope will open up new possibilities and will increase the sample size of spatially resolved observations. Studying models that produce suitable synthetic observables for those instruments will be especially useful for identifying the wind tracers that allow us to distinguish wind-driving mechanisms from observations. 
\section{Summary and conclusions}
\label{sec:conclusions}
In this work, we present photoevaporative disk-wind models to interpret atomic ($\mathrm{[OI]}\,  ^1\mathrm{D}_2\!-\!{^3}\mathrm{P}_2$ at $0.63\,\mathrm{\mu m}$) and molecular ($\mathrm{o\!-\!H_2}\,1\!-\!0\,\mathrm{S}(1)$ at $2.12\;\mathrm{\mu m}$) line emission from protoplanetary disks that show signatures of winds. The modelling approach consists in the post-processing of hydrodynamic X-ray photoevaporative disk-wind models, using the radiation--thermochemical model \prodimo to produce synthetic spectral line emission. 

We used this framework to model the observations of \oiline and \htline from \citet{Gangi2020}, including the line fluxes and the kinematic signatures. In this first attempt, we neglect the dynamics for the chemical calculations. Therefore, the models might under-predict the wind signatures (i.e. blueshift of the peak emission) of molecular hydrogen for cases with strong and dense winds. Nevertheless, we find that X-ray-driven photoevaporative wind models are generally consistent with the observed wind signatures for both the \oiline and \htline spectral lines in the currently available observational data.

We find blueshifted peaks in the \htline synthetic line profiles in the range of \mbox{$\mathrm{v_p}\!=\!0$ to $\!\approx\!-6\,\mathrm{km s^{-1}}$}. Also, the measured FWHMs of the modelled lines are consistent with the data.  However, we find that it is not required that both lines be emitted from the same regions close to the star, as suggested by \citet{Gangi2020} to explain the observationally derived kinematic wind signatures. In our models, the outer radius of the physical emitting regions for the \htline is always significantly larger than for \oiline, that is, by factors of a few to $\approx\!10$. The complex velocity field structure, the different vertical emitting layers, and the limited spectral resolution make the measured FWHM appear similar. Thus, we conclude that for currently available observational data, the approach of using the FWHM of the \oiline and \htline spectral line profiles and assuming only Keplerian broadening is not well suited to deriving the physical emitting radii of those two lines. Therefore, this method does not allow us to discriminate between magnetically driven and photo-evaporative disk winds. Further modelling of atomic and molecular line emission for both photoevaporative and magnetically driven winds is required in order to determine whether or not it is at all possible to determine the wind-driving mechanism from such spatially unresolved line emission. 

The observational data are still scarce and most targets in the sample of \cite{Gangi2020} do not show clear signatures of blueshifted emission in molecular hydrogen, which might be a consequence of the limited spectral resolution of the data. On the other hand, our models indicate that simply not enough molecular hydrogen can survive in low-density winds (such as photo-evaporative winds) in particular if the target has a high accretion luminosity, which results in efficient photo-dissociation of molecular hydrogen by FUV radiation. Consequently, any \htline emission becomes dominated by that from the high-density regions at the disk surface where the kinematics become dominated by Keplerian rotation.

Further improvements to the models are required. In particular, an efficient coupling of the dynamics, chemistry, and thermal balance will allow more accurate predictions of molecular line emission  to be produced. However, those models are computationally very expensive, meaning that a modelling approach such as the one presented here will still be required for a comprehensive exploration of the vast parameter space (e.g. disk structure, varying stellar properties, chemical networks). In particular, such models  will also be able to produce spatially resolved synthetic observables, which will be useful for exploring the capability of upcoming facilities such as the ELT.

\begin{acknowledgements}
The authors are grateful to the referee for a constructive and positive report that improved the paper. The authors want to thank M. Gangi and B. Nisini for useful discussions on their data and for providing some of it in digitized form. We acknowledge the support of the Deutsche Forschungsgemeinschaft (DFG, German Research Foundation) Research Unit ``Transition discs'' - 325594231. This research was supported by the Excellence Cluster ORIGINS which is funded by the Deutsche Forschungsgemeinschaft (DFG, German Research Foundation) under Germany's Excellence Strategy - EXC-2094 - 390783311. CHR is grateful for support from the Max Planck Society. This research has made use of NASA's Astrophysics Data System. This research made use of Astropy, a community-developed core Python package for Astronomy \citep{AstropyCollaboration2013,AstropyCollaboration2018}, matplotlib \citep{Hunter2007} and scipy \citep{2020SciPy-NMeth}.
\end{acknowledgements}
\bibliographystyle{aa}
\bibliography{h2diskwinds}
\begin{appendix} 
\section{Dust density structures}
\begin{figure}[ht]
    \centering
    \includegraphics[width=\hsize]{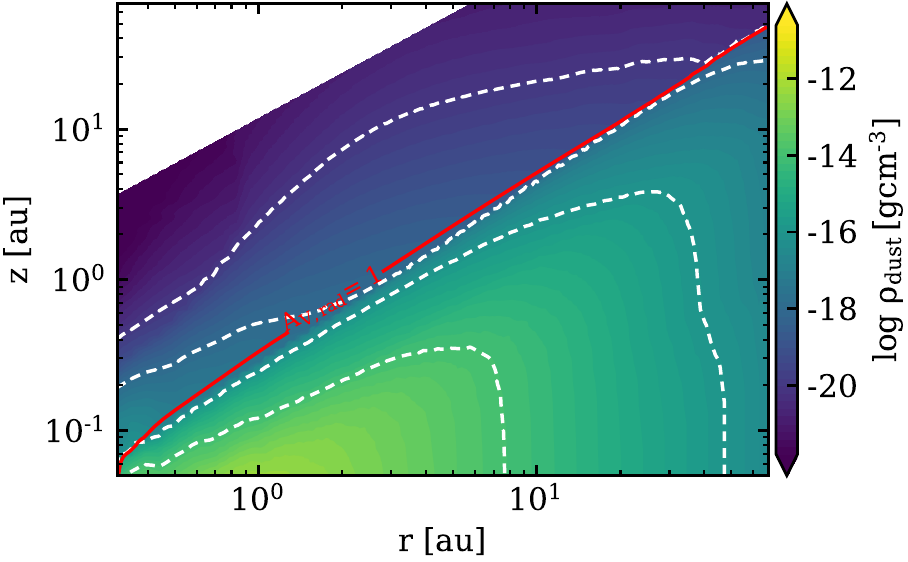}
    \includegraphics[width=\hsize]{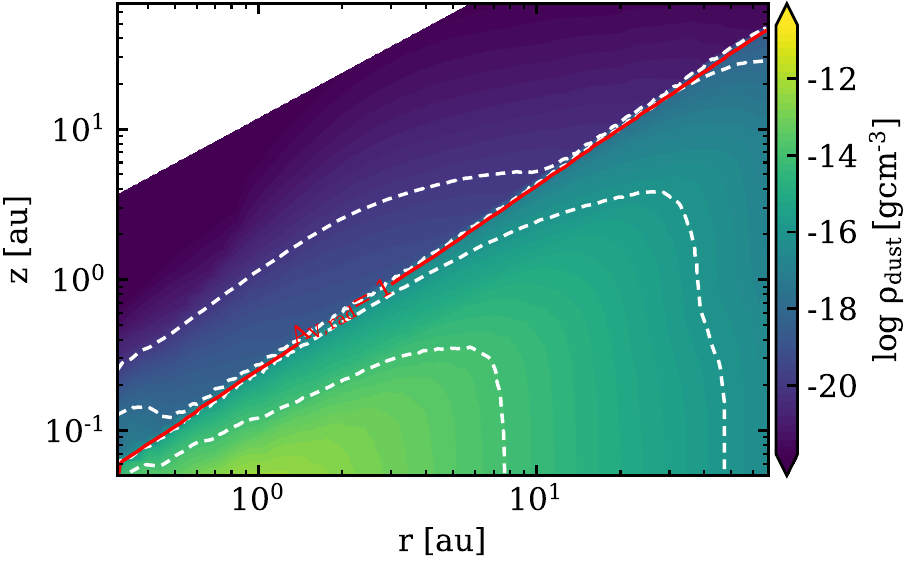}
    \caption{Dust density structure for the models with a constant gas-to-dust mass ratio of  $g/d=100$ in the disk and wind (top panel) and the model with a factor 10 lower dust density  $g/d=1000$ in the wind region (bottom panel). The red solid contour shows where the radial visual extinction $A_\mathrm{V,rad}$ equals unity. The white dashed contours correspond to the values shown in the colour bar.}
    \label{fig:dust}
\end{figure}

In Fig.~\ref{fig:dust} we show the dust density structure for our fiducial model and for the model where we lowered the dust density in the wind by a factor of 10 (gd models), resulting in a gas-to-dust mass ratio of 1000 (see Sect.~\ref{sec:metmodelseries}).

\section{Comparison to MOCASSIN}
\label{sec:moccomp}
Here we compare the results of the \prodimo model to the results of the MOCASSIN model of \citet{Weber2020}. We do not attempt to do a full benchmark test. Rather, we are  interested in how the two codes compare if the same physical input properties are used, such as the disk--wind density structure, stellar properties, and dust opacities. The two codes are conceptually different, in particular for the chemistry (i.e. \prodimo includes molecules) and the heating and cooling processes. It is therefore especially interesting to see the differences in the code for e.g. the temperature structure. 

\begin{figure*}
    \centering
    \includegraphics[width=\textwidth]{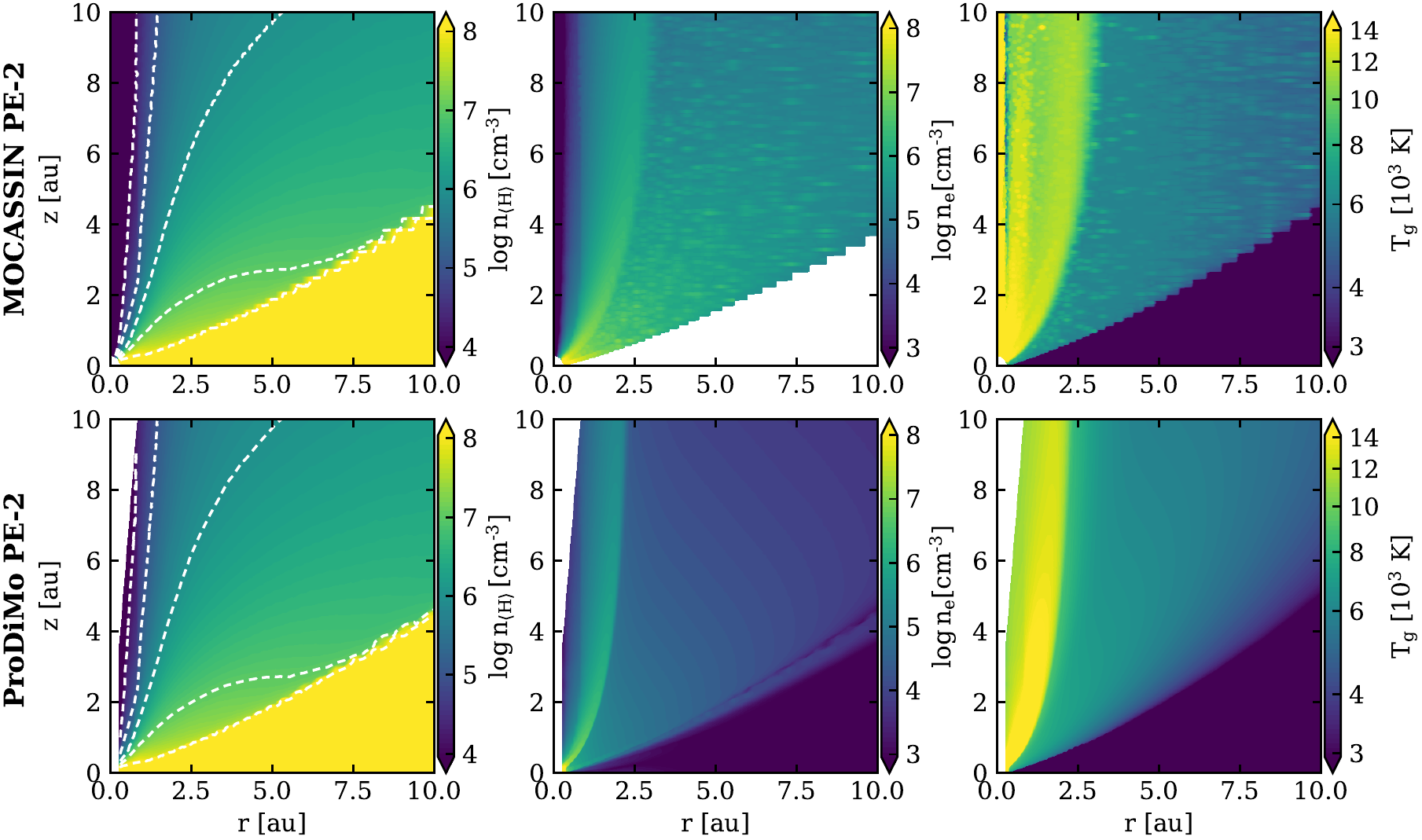}
    \caption{Comparison of the MOCASSIN and \prodimo model. The first row shows the MOCASSIN model, the bottom row the \prodimo model. From left to right the 2D structure of the hydrogen density (as reference, identical in both models), electron density and the gas temperature for the inner $\mathrm{10\,au}$ are shown.}
    \label{fig:cMOC_2D}
\end{figure*}

\begin{figure*}
    \centering
    \includegraphics[width=\textwidth]{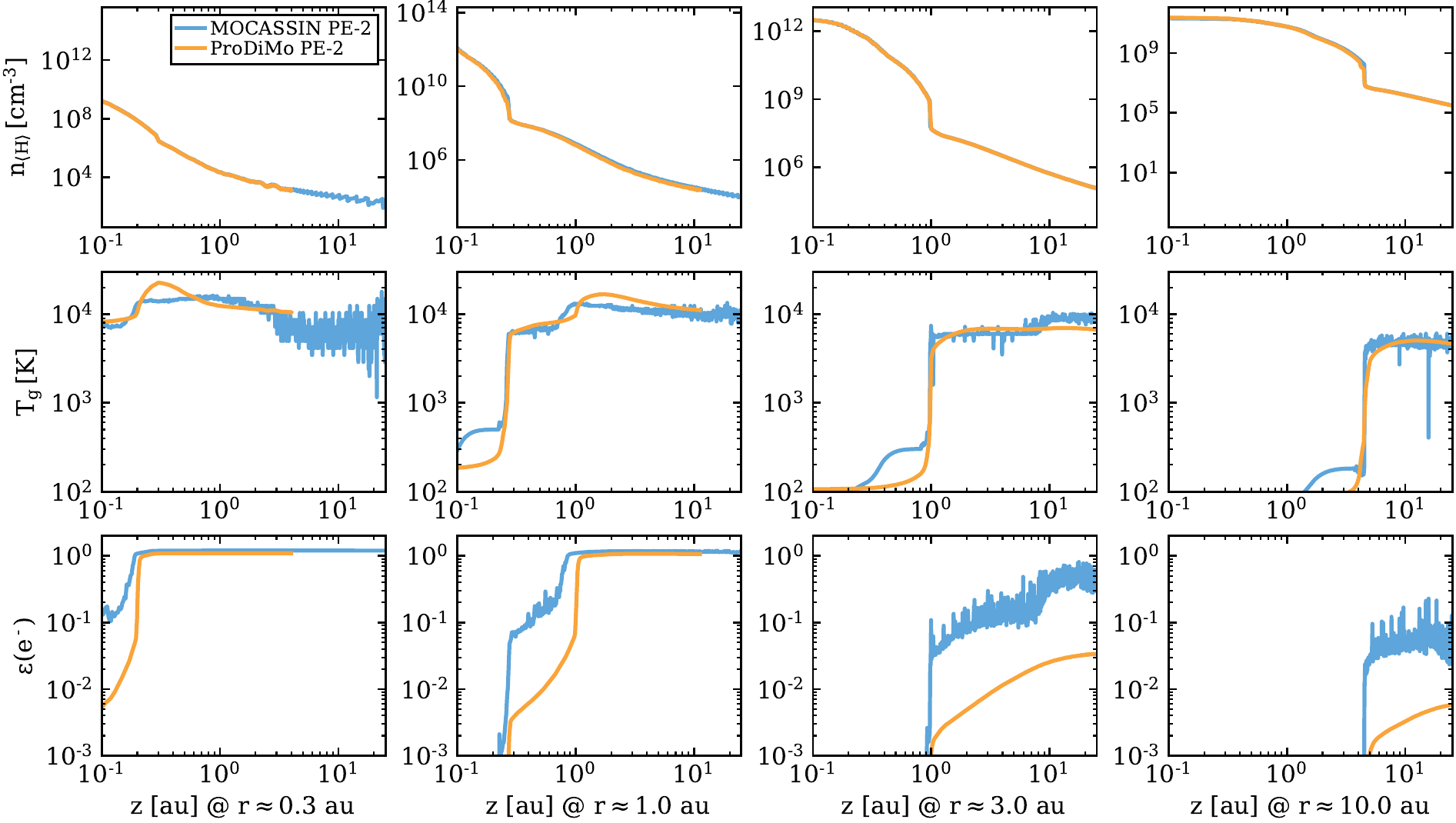}
    \caption{Comparison of the \prodimo (orange colour) and the MOCASSIN (blue colour) model for vertical cuts through the disk at $r\approx0.3,1,3$ and $10\,\mathrm{au}$ (columns from left to right). The rows (top to bottom) show the comparison for the hydrogen number density (for reference, identical in the two models), gas temperature, and electron abundance (relative to the total hydrogen density), respectively.}
    \label{fig:cMOC_ver}
\end{figure*}

Fig.~\ref{fig:cMOC_2D} shows a comparison for the electron number density $\mathrm{n_e}$ (ionisation degree) and the gas temperature $\mathrm{T_g}$ for the 2D structure. Fig.~\ref{fig:cMOC_ver} shows the comparison for vertical cuts through the disk structure. These figures show that the general features in $\mathrm{n_e}$ and $\mathrm{T_g}$ are very similar. However, there are also differences. The regions with $\mathrm{T_g}\!>\!10000\,\mathrm{K}$ are more extended in the MOCASSIN model. This is likely caused by differences in the X-ray radiative transfer. MOCASSIN places the EUV/X-ray source above and below the star ($\approx 10\,\mathrm{R_*})$ whereas in \prodimo the EUV/X-ray source is at the position of the star (for technical reasons). Placing the emitting source higher above the disk midplane allows the radiation to penetrate deeper into the disk/wind. This will also affect $\mathrm{n_e}$ in the wind due to direct ionisation of hydrogen. However, those effects should not be significant and they only affect low-density regions \citep{Ercolano2013b}. Otherwise, the gas temperatures agree within a factor of two in the disk wind region (i.e. high up in the disk; see also Fig.~\ref{fig:cMOC_ver}). At the transition from the wind region to the disk the differences in $\mathrm{T_g}$ become significant, which is expected as \mbox{MOCASSIN} does not include molecules, which can become important coolants in that region. 

For the electron density (abundance), the differences are more significant (about a factor of 10), in particular in the outer disk ($r\gtrsim 3\,\mathrm{au}$). This is somewhat expected, because \prodimo includes molecules, which affect the ionisation chemistry. Furthermore, differences in wavelength-dependent ionisation cross-sections likely also play a role in addition to the differences in the radiative transfer methods.

Considering that \prodimo and MOCASSIN are conceptually different codes, the results compare reasonably well. We therefore conclude that using \prodimo in combination with radiation-hydrodynamic photoevaporative wind models is a suitable approach that also allows molecular line emission to be modelled, in particular molecular hydrogen. 

\clearpage

\section{Physical properties in the line-emitting regions}
\label{sec:physprops}
Table~\ref{table:lineorig}  summarises certain physical properties of the line-emitting regions shown in Fig.~\ref{fig:lineorig}. To calculate those quantities, we simply average the respective quantity over the whole emitting region. From an observational perspective, those values are only strictly valid for a disk viewed face-on, as not
all emission is seen depending on the disk inclination. Nevertheless, those values are representative of the physical conditions for line excitation. We note that a mass-weighted average does affect the absolute numbers, but the general picture and trends remain the same.

\begin{table}[h!]
\label{table:lineorig} 
\setlength{\tabcolsep}{3pt}
\caption{Averaged quantities of the line-emitting regions shown in Fig.~\ref{fig:lineorig}.} 
\resizebox{\hsize}{!}{
\begin{tabular}{llrrcccr}
\hline\hline
Model & Line & $T_\mathrm{gas}$ & $T_\mathrm{dust}$ & \tablefootmark{a}$n_\mathrm{e^-}$ & \tablefootmark{b}$n_\mathrm{H}$ & \tablefootmark{c}$n_\mathrm{tot}$ & \tablefootmark{d}$\varv_\mathrm{wind}$ \\
&  & [K] & [K] & [$\mathrm{cm^{-3 }}$] & [$\mathrm{cm^{-3 }}$] & [$\mathrm{cm^{-3 }}$] & [$\mathrm{km s^{-1 }}$] \\
\hline
PE-1    & [OI]    & 8050 & 202 & 1.2(5)\tablefootmark{e} & 4.8(6) & 4.9(6) & 5.40 \\
        & o-H$_2$ &  539 &  95 & 2.3(2) & 2.8(5) & 7.3(5) & 13.27 \\
PE-2    & [OI]    & 6831 & 221 & 6.9(4) & 9.5(6) & 9.6(6) &  6.69 \\
        & o-H$_2$ & 1227 &  96 & 2.6(2) & 5.5(5) & 9.7(5) &  8.81 \\
PE-3    & [OI]    & 6792 & 229 & 5.1(4) & 1.2(7) & 1.2(7) &  6.71 \\
        & o-H$_2$ & 1088 & 104 & 3.1(2) & 9.3(5) & 3.1(6) &  3.19 \\
\hline                                             
PE-1 gd & [OI]    & 7724 & 212 & 1.5(5) & 8.8(6) & 8.9(6) &  4.64 \\
        & o-H$_2$ &  223 &  93 & 3.6(2) & 7.8(5) & 1.2(6) &  7.79 \\
PE-2 gd & [OI]    & 4582 & 206 & 5.5(4) & 2.9(7) & 3.0(7) &  3.66 \\
        & o-H$_2$ &  362 &  92 & 6.7(2) & 2.4(6) & 4.1(6) &  4.17 \\
PE-3 gd & [OI]    & 2706 & 159 & 2.3(4) & 4.8(7) & 5.2(7) &  1.10 \\
        & o-H$_2$ & 1095 & 104 & 2.6(3) & 1.3(7) & 1.9(7) &  0.59 \\
\hline
\end{tabular}}
\tablefoot{
\tablefoottext{a}{electron number density;}
\tablefoottext{b}{number density of neutral hydrogen;}
\tablefoottext{c}{total number density;}
\tablefoottext{d}{$\varv_\mathrm{wind}=\sqrt{{\varv_r}^2+{\varv_z}^2}$;}
\tablefoottext{e}{$a(b)$ means $a\times10^b$}
}
\end{table}
\section{Line luminosities from the literature}
\label{sec:litlum}
In addition to the data from \citet{Gangi2020}, we also collected data from the literature for targets with observations for both the \oiline and \htline lines. The collected line luminosities and the adapted distances to calculate them are listed in Table~\ref{table:litlum}. We note that for this work we only use the luminosities from the literature data as the kinematic data for the \htline were often not available (i.e. due to low spectral resolution). 

From the collected targets, only GM Aur is also included in \citet{Gangi2020}, and these authors report $0.18\pm0.02\times10^{-5}\,L_\mathrm{\odot}$ and $0.4\pm0.1\times10^{-5}\,L_\mathrm{\odot}$, for the \htline and the \oiline line, respectively. Those values are a factor 4.6 higher for \htline and a factor 4.2 lower for  \oiline, compared to the values listed in Table~\ref{table:litlum}. The origin of those differences are unclear, but might be related to variability (although this would be rather strong, see \citealt{Gangi2020}), different definitions of the NLVC for \oiline or unknown systematic errors (i.e. observations were done with different instruments).  
\begin{table}[h!]
\caption{Collected line luminosities from the literature.}
\label{table:litlum}
\centering
\begin{tabular}{lccc}
\hline\hline
Target & \oiline & \htline & distance \\
 &       $[10^{-5}\,L_\odot]$  & $[10^{-5}\,L_\odot]$  & [pc]\\
\hline 
TW Hya   & $~~0.933\pm0.187$\tablefootmark{1} & $0.011\pm0.001$ \tablefootmark{2,3}&  ~59.8 \\
LkCa 15  & $~~0.871\pm0.174$\tablefootmark{1} & $0.131\pm0.015$\tablefootmark{2} & 156.9 \\
GM Aur   &  $~~1.700\pm0.170$\tablefootmark{4} & $0.039\pm0.001$ \tablefootmark{5} & 159.0 \\
AA Tau   &  $~~0.468\pm0.047$\tablefootmark{4} & $0.196\pm0.006$\tablefootmark{5} & 140.0 \\
V773 Tau & $13.804\pm1.380$\tablefootmark{4} & $0.139\pm0.005$ \tablefootmark{5} & 131.0 \\
CS Cha   & $~~2.240\pm1.200$\tablefootmark{6} & $0.168\pm0.008$\tablefootmark{7} & 160.0 \\
\hline
\end{tabular}
\tablefoot{
References: \tablefoottext{1}{\citet{Fang2018}}. \tablefoottext{2}{\citet{Bary2003}}; \tablefoottext{3}{\citet{Nomura2005b}}; \tablefoottext{4}{\citet{Simon2016}};  \tablefoottext{5}{\mbox{\citet{Beck2019}};} \tablefoottext{6}{\citet{Manara2014c}}; \tablefoottext{7}{\citet{Bary2008}}}
\end{table}

\newpage
\section{Selected targets for comparison to the models}
\label{sec:targetscomp}
Table~\ref{table:targetscomp} summarises the kinematic properties of the targets used for the comparison of the measured line kinematics to the models. All those targets have a blueshifted NLVC in both \oiline and \htline. All data listed in Table~\ref{table:targetscomp} are from \citet{Gangi2020}.
\begin{table}[h!]
\caption{Targets used for the comparison of the line kinematics to the models.}
\label{table:targetscomp}
\setlength{\tabcolsep}{3pt}
\resizebox{\hsize}{!}{\begin{tabular}{lcrrccc}
\hline\hline
Target & i  & [OI] $\varv_\mathrm{p}$\hspace*{0.1cm} & $\mathrm{H_2}$ $\varv_\mathrm{p}$\hspace*{0.2cm} & \parbox[c]{1cm}{\centering[OI] FWHM\vspace*{0.05cm}} & \parbox[c]{1cm}{\centering$\mathrm{H_2}$ FWHM\vspace*{0.05cm}} & \parbox[c]{1.3cm}{\centering [OI]\tablefootmark{1}\\other\hspace*{0.1cm}}\\
       & [$^\circ$] & [$\mathrm{km s^{-1}}$]  & [$\mathrm{km s^{-1}}$] & [$\mathrm{km s^{-1}}$] & [$\mathrm{km s^{-1}}$] & comp.\\
\hline 
DG Tau  &  37.0  & $-6.8\!\pm\!0.4 $  & $-7.8\!\pm\!1.0 $  & $ 16.3\!\pm\!0.2 $ & $14.2\!\pm\!1.3 $  & HVC\\
DO Tau  &  27.6  & $-14.3\!\pm\!3.2 $ & $-15.7\!\pm\!2.1 $ & $ 24.8\!\pm\!7.7 $ & $11.1\!\pm\!1.5 $ & HVC\\
GM Aur  &  53.2  & $-0.9\!\pm\!3.9 $  & $-1.4\!\pm\!1.7 $  & $ 11.0\!\pm\!4.9 $ & $17.6\!\pm\!2.8 $  & None\\
HN Tau  &  69.8  & $-7.5\!\pm\!1.8 $  & $-5.3\!\pm\!3.4 $  & $ 47.6\!\pm\!1.3 $ & $37.6\!\pm\!1.8 $  & unclear\tablefootmark{2}\\
UX Tau  &  39.0  & $-2.0\!\pm\!0.5 $  & $-6.0\!\pm\!1.1 $  & $ 27.7\!\pm\!2.2 $ & $11.2\!\pm\!2.7 $  & \parbox[c]{1.3cm}{\centering redshift \\BLVC\vspace*{0.1cm}}\\
UY Aur  &  23.5  & $-0.5\!\pm\!1.9 $  & $-0.3\!\pm\!1.7 $  & $ 18.2\!\pm\!3.4 $ & $12.7\!\pm\!0.5 $  & BLVC\\
XZ Tau  &  35.0  & $-0.9\!\pm\!1.9 $  & $-5.8\!\pm\!1.8 $  & $ 13.1\!\pm\!3.8 $ & $16.9\!\pm\!0.5 $  & unclear\\
\hline
\end{tabular}}
\tablefoot{\tablefoottext{1}{This column indicates if the observational data show other kinematic components besides the NLVC for  \oiline according to the Gaussian fits of \citet{Gangi2020}.} \tablefoottext{2}{\textit{unclear} means that there is emission at high velocities but it was not possible to fit it with a Gaussian.}}
\end{table}
\section{Gaussian fitting of line profiles}
\label{sec:linefit}
\begin{figure*}
    \includegraphics[width=\textwidth]{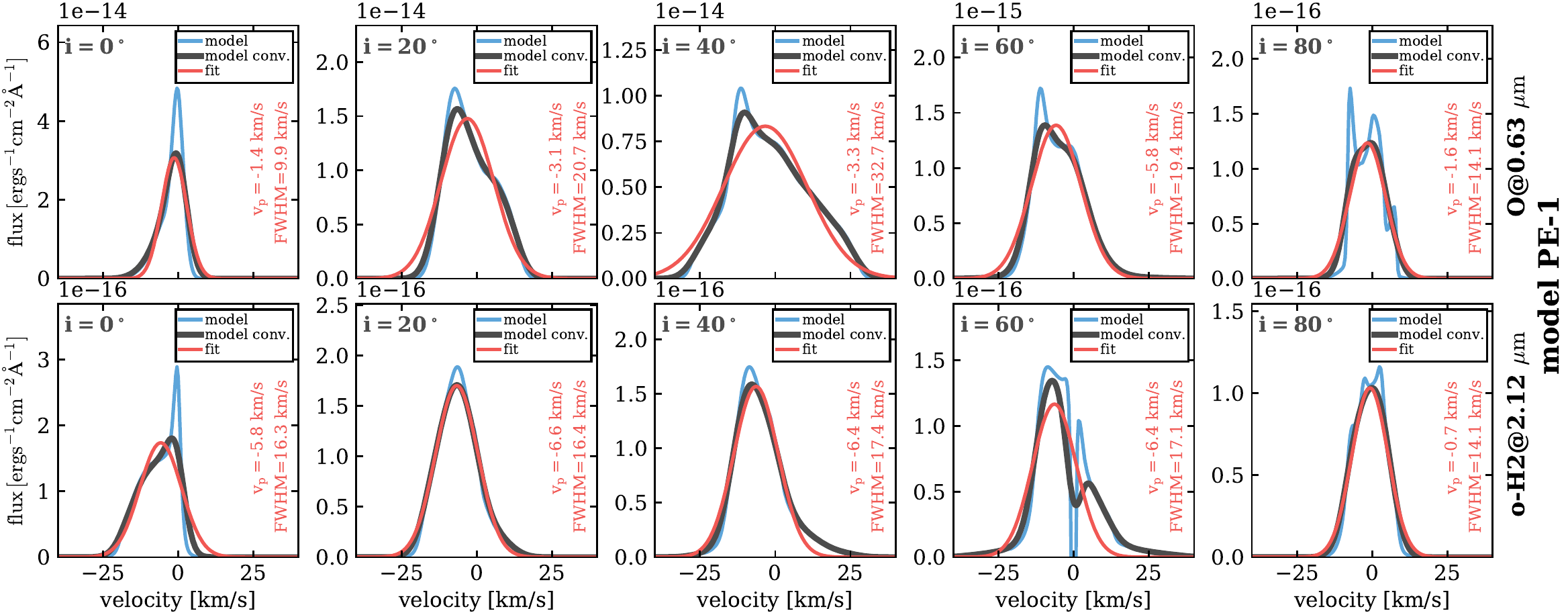}
    
    \includegraphics[width=\textwidth]{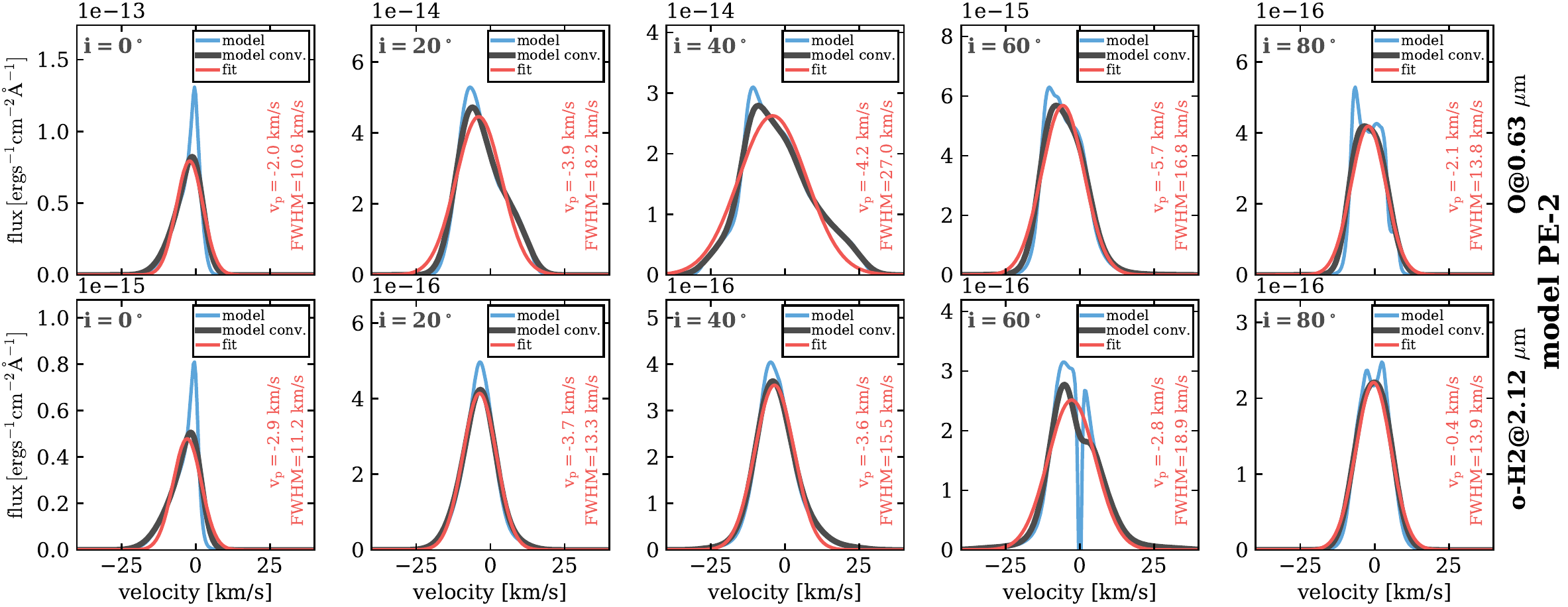}
    
    \includegraphics[width=\textwidth]{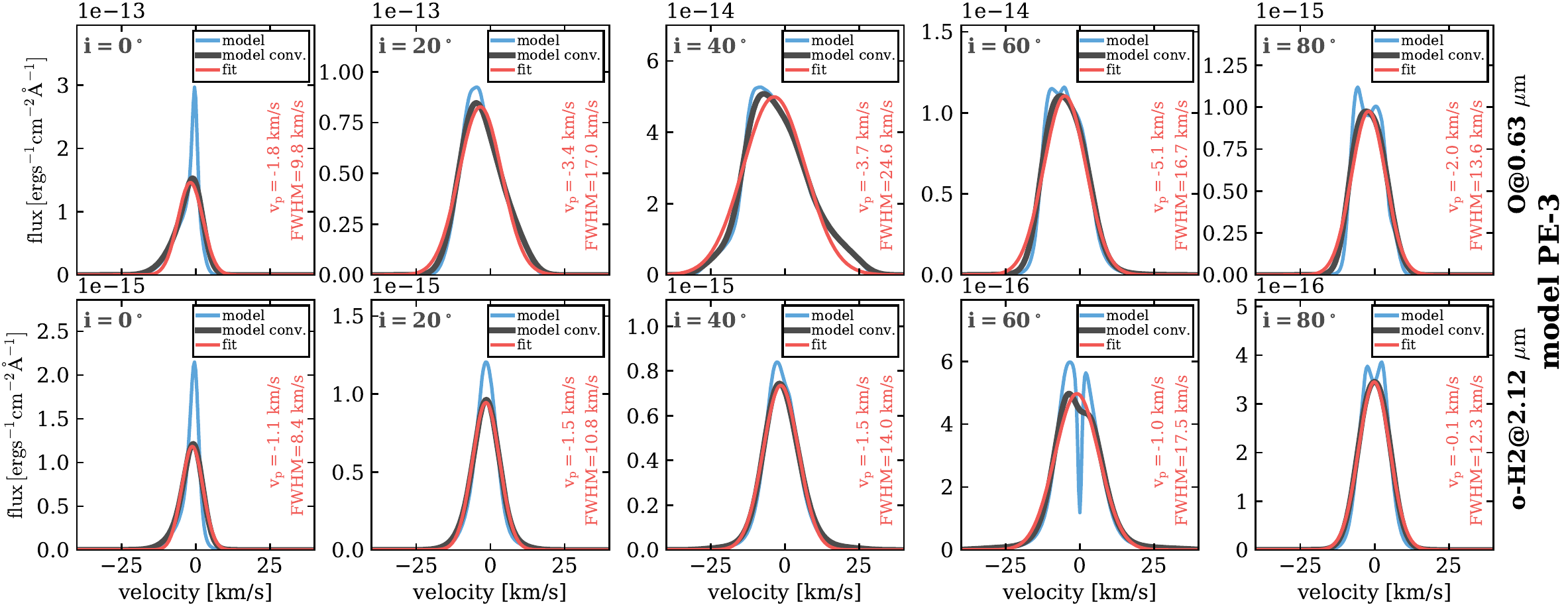}    
    \caption{Fitting of the modelled line profiles. From top to bottom, the models PE-1, PE-2, and PE-3 are shown. Please note the different scaling on the y axis (flux) for each panel. For each model, we show the \oiline (first row) and the \htline  (second row) for five different inclinations (columns from left to right). Each panel shows the line profile with the model resolution (blue) convolved to a resolution of $R=50000$ similar to the observations (black), and the fit (the NLVC) to the convolved profile (red solid line).}
\label{fig:linefit}
\end{figure*}

\begin{figure*}
    \includegraphics[width=\textwidth]{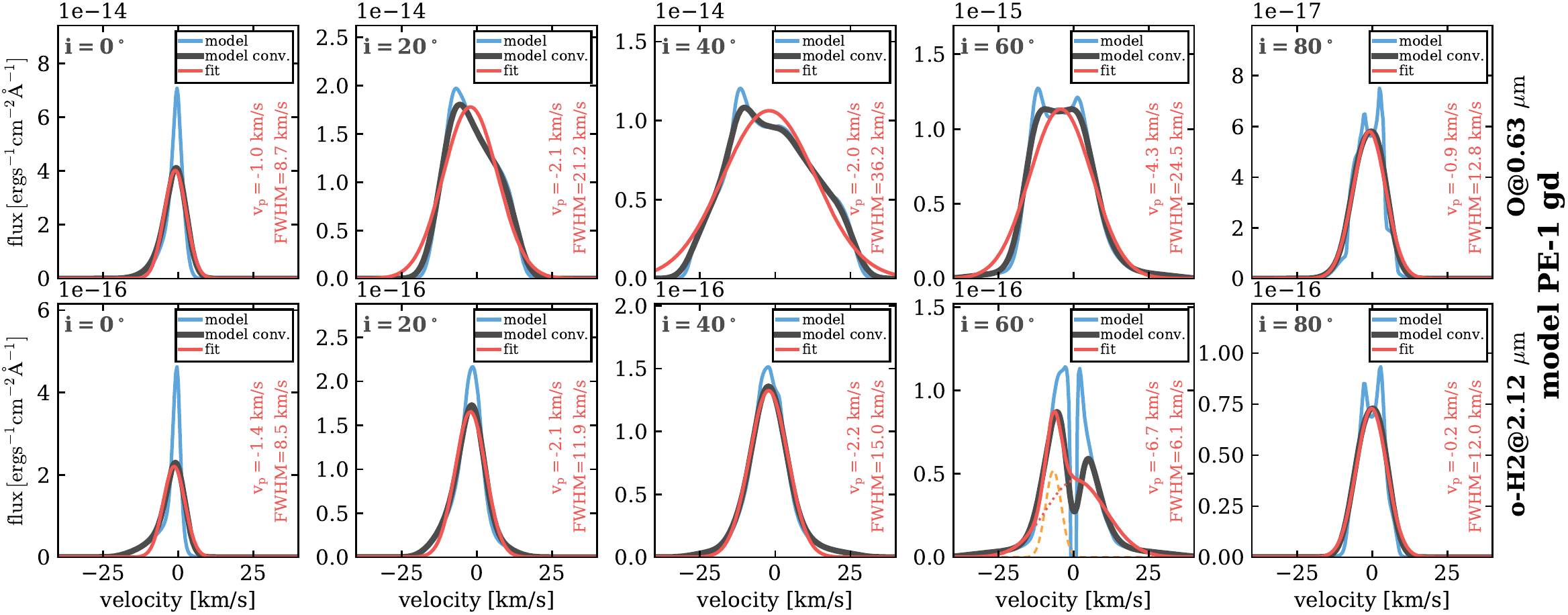}
    
    \includegraphics[width=\textwidth]{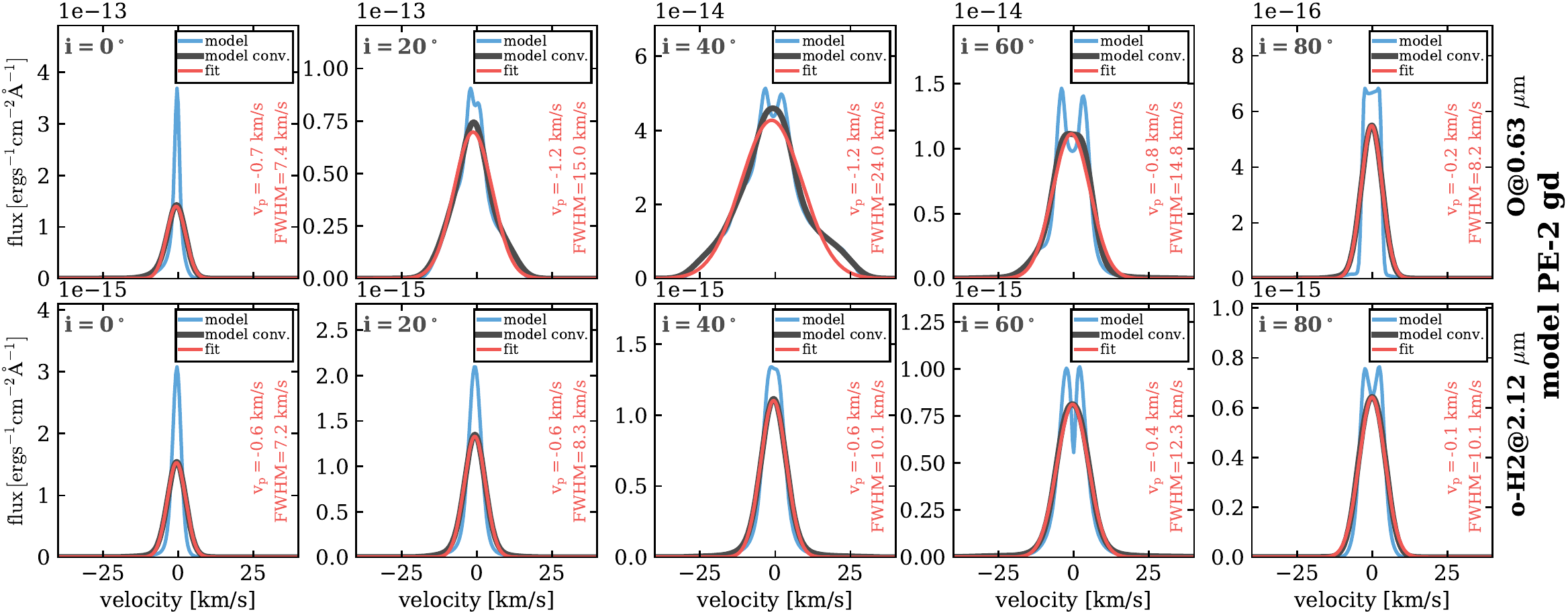}

    \includegraphics[width=\textwidth]{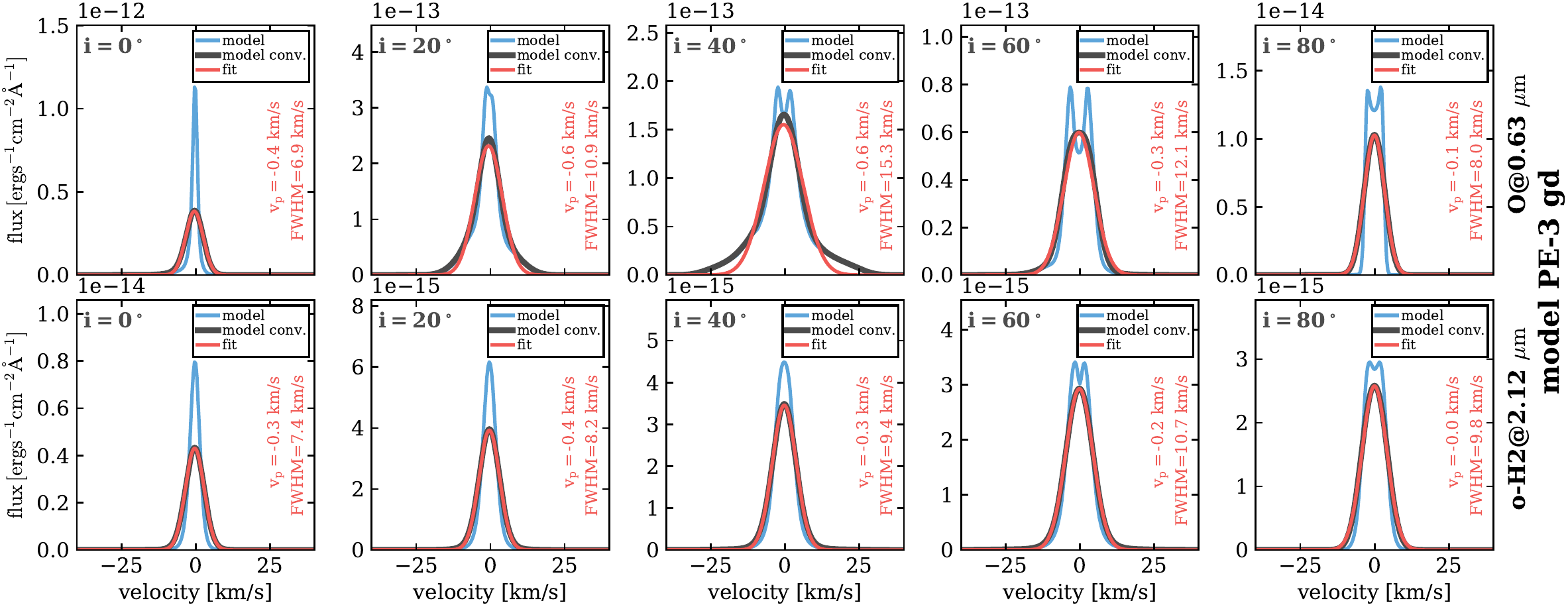}    
    \caption{Same as Fig.~\ref{fig:linefit} but for the models with a factor of ten lower dust density in the wind. The red and orange dotted lines show the individual components of the fit (not seen in the case of a single Gaussian). In the case of multiple Gaussians, the orange dotted line indicates the one chosen as the NLVC used to determine $\varv_\mathrm{p}$ and the FWHM.}
\label{fig:linefitgd}
\end{figure*}

Figs.~\ref{fig:linefit} and~\ref{fig:linefitgd} show the results of the Gaussian fitting process (see Sect.~\ref{sec:metlines}) for all photoevaporative models presented in this work using the observational spectral resolution of $R\approx 50000$. 

\begin{figure*}
    \includegraphics[width=\textwidth]{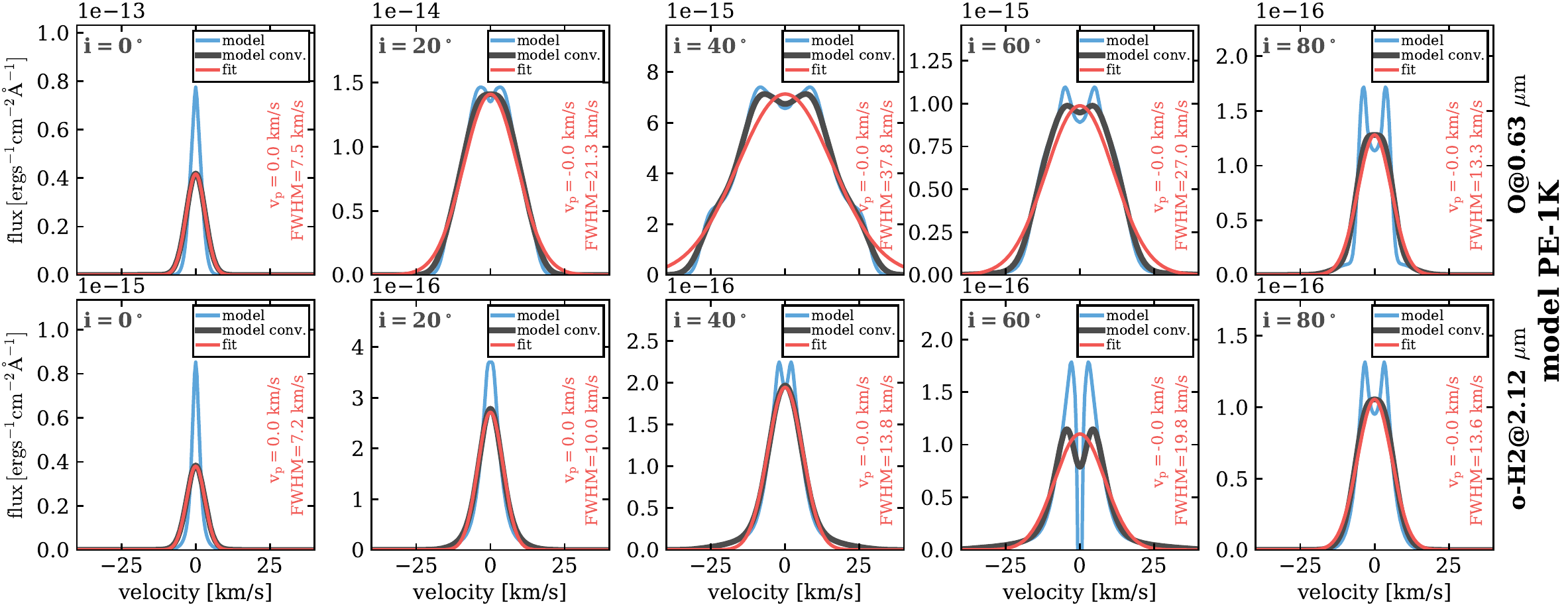}
    
    \includegraphics[width=\textwidth]{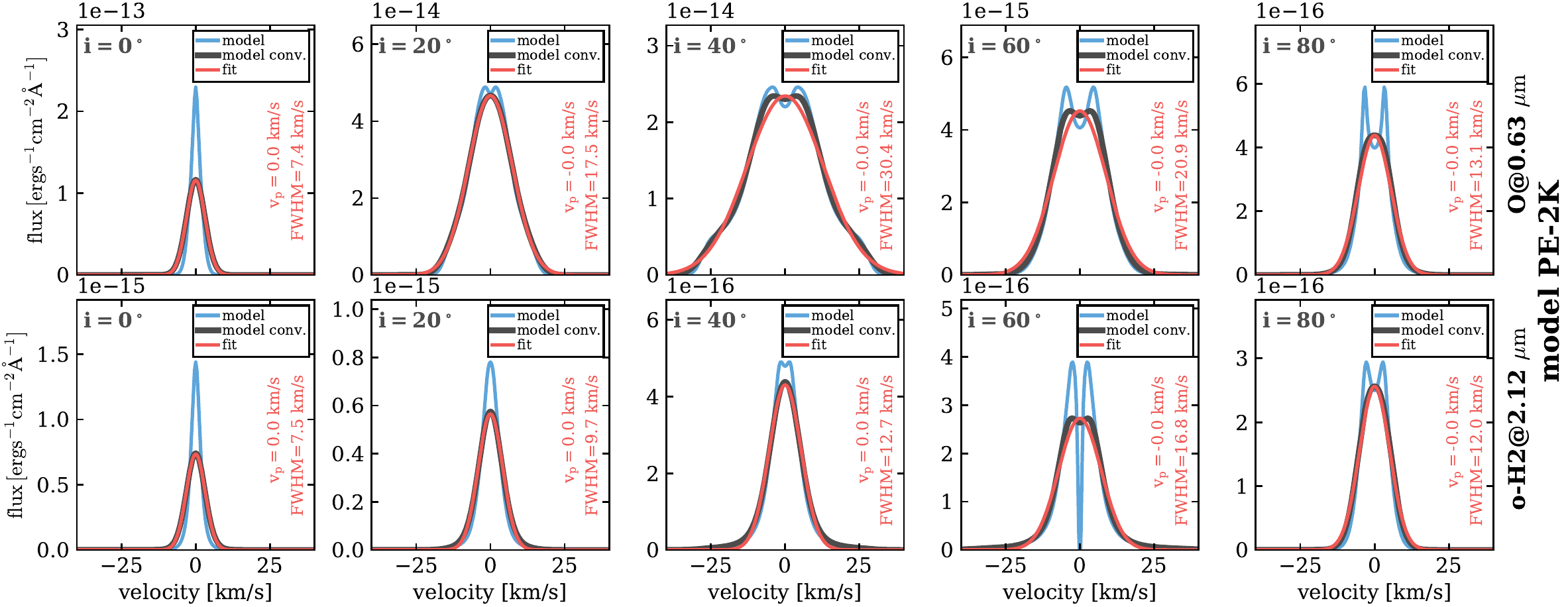}
    
    \includegraphics[width=\textwidth]{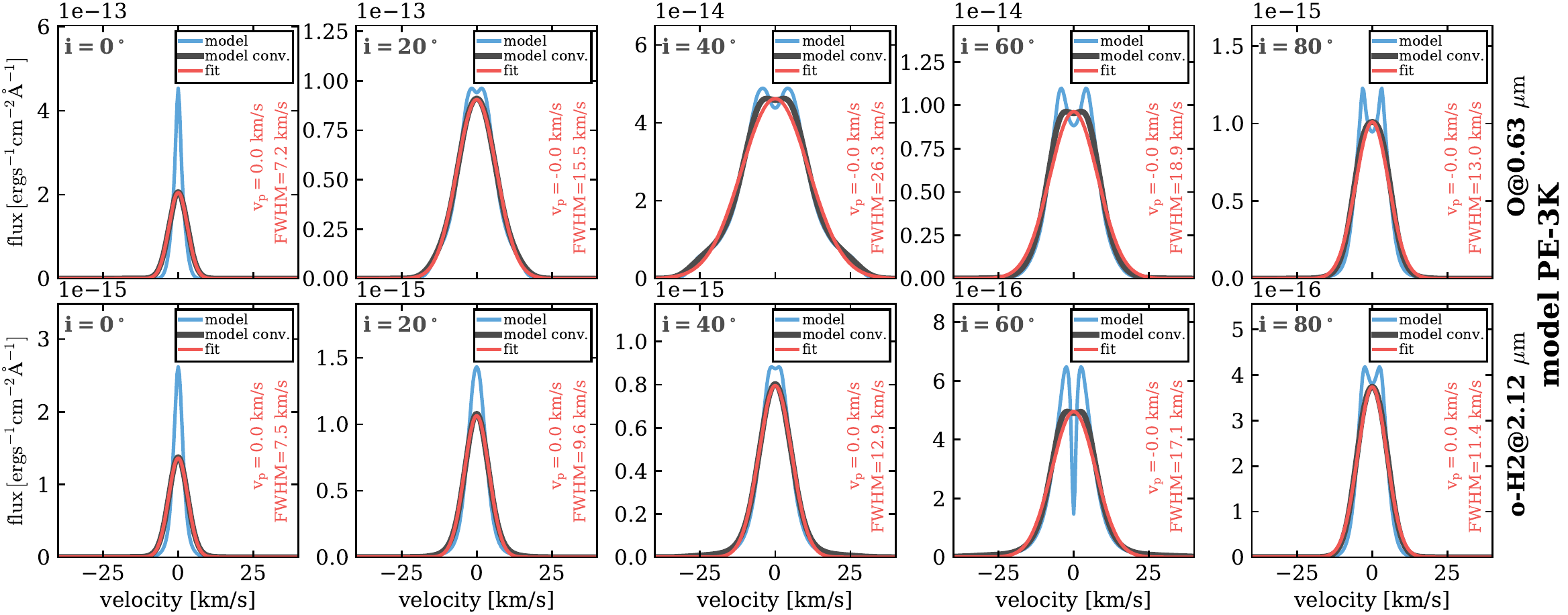}       
    \caption{Same as Fig.~\ref{fig:linefit} for the PE-1, PE-2, and PE-3 models, but assuming a purely Keplerian velocity field (see Sect.~\ref{sec:disthindisk} for details).}
\label{fig:linefitK}
\end{figure*}

Fig.~\ref{fig:linefitK}  shows the fitting results of the modelled profiles assuming a purely Keplerian velocity field. As opposed to the models with the wind velocity field, these latter profiles are always symmetric and hence the fitting routine always gives $\varv_\mathrm{p}=0\,\mathrm{kms^{-1}}$. The impact of Keplerian assumption concerning the FWHM is discussed in Sect.~\ref{sec:disthindisk}.
\end{appendix}
\end{document}